\documentclass{aa}

\usepackage{graphicx}
\usepackage{txfonts}
\usepackage{hyperref}
\graphicspath{{Images/}{../Images/}}
\usepackage[normalem]{ulem}
\usepackage{aas_macros}

\begin{document}

    \title{Jupiter’s interior from Juno:\\ Equation-of-state uncertainties and dilute core extent\thanks{Models of  Tables E.1. and E.2. are available at the CDS via anonymous ftp to \href{http://cdsarc.cds.unistra.fr}{cdsarc.cds.unistra.fr} (\href{ftp://130.79.128.5/}{130.79.128.5})
or via \url{https://cdsarc.cds.unistra.fr/cgi-bin/qcat?J/A+A/} and at \url{https://doi.org/10.5281/zenodo.7598377}.}}


    \author{S. Howard \inst{1}
          \and T. Guillot\inst{1}
          \and M. Bazot\inst{2,3}
          \and Y. Miguel\inst{4,5}
          \and D. J. Stevenson\inst{6}
          \and E. Galanti\inst{7}
          \and Y. Kaspi\inst{7}
          \and W. B. Hubbard\inst{8}
          \and B. Militzer\inst{9}
          \and R. Helled\inst{10}
          \and N. Nettelmann\inst{11,12}
          \and B. Idini\inst{6}
          \and S. Bolton\inst{13}
          }

    \institute{Université Côte d'Azur, Observatoire de la Côte d'Azur, CNRS, Laboratoire Lagrange, France\\
              \email{saburo.howard@oca.eu}
         \and
             Heidelberg Institute for Theoretical Studies (HITS gGmbH), Schloss-Wolfsbrunnenweg 35,69118 Heidelberg, Germany              
         \and
             CITIES, NYUAD Institute, New York University Abu Dhabi, PO Box 129188, Abu Dhabi, United Arab Emirates,
         \and
             SRON Netherlands Institute for Space Research , Niels Bohrweg 4, 2333 CA Leiden, the Netherlands,
         \and
             Leiden Observatory, University of Leiden, Niels Bohrweg 2, 2333CA Leiden, The Netherlands,
         \and
             Division of Geological and Planetary Sciences, California Institute of Technology, Pasadena, California 91125, USA,
         \and
             Department of Earth and Planetary Sciences,Weizmann Institute of Science, Rehovot 76100, Israel,
         \and
             Lunar and Planetary Laboratory, University of Arizona, Tucson, AZ 85721, USA,
         \and
             Department of Earth and Planetary Science, University of California, Berkeley, CA 94720, USA,   
         \and
             Institute for Computational Science, University of Zurich, Winterthurerstr. 190, CH8057 Zurich, Switzerland,
         \and
             Department of Astronomy and Astrophysics, University of California, Santa Cruz, CA 95064, USA
         \and
             Institute of Planetary Research, German Aerospace Center, D-12489 Berlin, Germany
         \and
             Southwest Research Institute, San Antonio, TX 78238, USA   
             }

    \date{Accepted February 6, 2023}
    
 
  \abstract
   {The Juno mission has provided measurements of Jupiter's gravity field with an outstanding level of accuracy, leading to better constraints on the interior of the planet. Improving our knowledge of the internal structure of Jupiter is key to understanding its formation and evolution but is also important in the framework of exoplanet exploration.}
   {In this study, we investigated the differences between the state-of-the-art equations of state and their impact on the properties of interior models. Accounting for uncertainty on the hydrogen and helium equation of state, we assessed the span of the interior features of Jupiter.}
   {We carried out an extensive exploration of the parameter space and studied a wide range of interior models using Markov chain Monte Carlo (MCMC) simulations. To consider the uncertainty on the equation of state, we allowed for modifications of the equation of state in our calculations.}
   {Our models harbour a dilute core and indicate that Jupiter's internal entropy is higher than what is usually assumed from the Galileo probe measurements. We obtain solutions with extended dilute cores, but contrary to other recent interior models of Jupiter, we also obtain models with small dilute cores. The dilute cores in such solutions extend to $\sim$~20\% of Jupiter's mass, leading to better agreement with formation--evolution models.} 
   {We conclude that the equations of state used in Jupiter models have a crucial effect on the inferred structure and composition. Further explorations of the behaviour of hydrogen--helium mixtures at the pressure and temperature conditions in Jupiter will help to constrain the interior of the planet, and therefore its origin.}


   \maketitle
%
\section{Introduction}

Despite the significant improvement to measured gravitational moments provided by  Juno \citep{2018Natur.555..220I,2020GeoRL..4786572D}, the interior of Jupiter remains mysterious. After the two first perijoves of Juno, \citet{2017GeoRL..44.4649W} proposed the presence of a dilute core inside the planet: a region above the central compact core where heavy elements are gradually mixed with hydrogen and helium in the envelope. However, most models led to atmospheric abundances that are incompatible with observations. \citet{2019ApJ...872..100D} then looked for models compatible with atmospheric abundances and proposed models that require an inward decrease of the abundance of heavy elements (negative Z gradient). While this cannot be ruled out, it seems unlikely from the point of view of long-term stability and formation models. Recently, \citet{2022PSJ.....3..185M} found models with both an atmosphere of protosolar composition of heavy elements and a positive Z gradient, but with large deviations of the gravitational moments requiring specific differential rotation solutions. In \citet{2022A&A...662A..18M}, we found solutions with smaller differential rotation offsets but these required a higher interior entropy than that measured by the Galileo probe.

While all these recent interior models rely on different assumptions, most of them still yield a dilute core inside Jupiter that is very extended. \citet{2019ApJ...872..100D} and \citet{2022PSJ.....3..185M} find dilute cores that respectively reach 65\%-75\% and 63\% of Jupiter's radius. These values correspond to a dilute core that extends up to respectively $\sim 60\%-75\%$ and $\sim 50\%$ of Jupiter's mass. From the point of view of evolution, \citet{2018A&A...610L..14V} showed that the outer envelope is mixing efficiently and the outer 60\% in mass are of uniform composition after 4.5~Gyr. However, \citet{2020A&A...638A.121M} then modelled the formation of Jupiter using realistic initial entropies and found even more efficient mixing during the planetary evolution where only the inner 20\% of the mass is left intact, suggesting that Jupiter's dilute core is not very extended. It has therefore been challenging so far to find agreement over the extent of the dilute core between interior and formation--evolution models of Jupiter, unless an additional process, such as a giant impact, is considered \citep{2019Natur.572..355L}. In addition,  the gravitational imprint of the dilute core in Jupiter's tidal signal registered by Juno \citep{2022PSJ.....3...89I,2022PSJ.....3...11I} has lead to increased uncertainty over the extent of the dilute core. \\

Furthermore, constraining the internal structure of Jupiter requires a good understanding of the behaviour of hydrogen and helium at the pressure and temperature conditions in the planet. Interior models therefore also rely on a key ingredient, which is a hydrogen and helium equation of state (hereafter H-He EOS). Experiments and simulations have been extensively conducted to provide accurate EOSs (see \citet{2020NatRP...2..562H} and references therein for a review) but some uncertainty remains. In addition, \citet{HowardGuillot2022} recently emphasised the importance of accounting for H-He interactions when calculating EOSs for interior models. The aim of this study is to estimate how extended Jupiter's dilute core is given the current uncertainty on the H-He EOS. \\

In Section~\ref{section:mcmc}, we first explain our methods and the details of our interior models. In Section~\ref{section:eos}, we compare some H-He EOSs in order to assess the uncertainty that must be accounted for in interior models. Section~\ref{section:results} is devoted to the results, where we present models with original EOSs but also models including a modification of the EOS.

\section{Methods}
  \label{section:mcmc}

Following the method described in \citet{2022A&A...662A..18M} \citep[see also][]{2018Natur.555..227G}, our approach is designed to obtain statistically robust estimates of the properties of Jupiter's interior.

\subsection{Observed physical characteristics}
\label{subsec:observable_data}

Interior models satisfy Jupiter's mass, which is obtained through $GM = 1.266865341 \times 10^{17}$~$\rm m^3/s^2$ \citep{2020GeoRL..4786572D}. Jupiter's equatorial radius has also been measured $R_{\rm eq}=71492 \pm 4$~km \citep{1992AJ....103..967L}. Jupiter's fast rotation, in 9hr55min29.7s \citep{1986CeMec..39..103D} flattens the planet and the departure from sphericity can be measured through the gravitational moments, which themselves are measured very accurately by Juno \citep{2020GeoRL..4786572D}:
\begin{equation}
    J_{2n}=-\frac{1}{MR_{\rm eq}^{2n}}\int \rho(r')^{2n}P_{2n}(cos \theta)d^3r'
,\end{equation}
where $R_{\rm eq}$ is Jupiter's equatorial radius, $\rho$ the density, $P_{2n}$ the Legendre polynomials, and $r$ and $\theta$ the radius and the colatitude, respectively. We aim to find models in agreement with the observed equatorial radius and gravitational moments. As Jupiter exhibits zonal winds in its upper region, we account for the contribution of the latitude-dependent differential rotation to the even gravitational moments \citep{2017GeoRL..44.5960K,2018Natur.555..223K,2018Natur.555..227G}. Our interior models must therefore match $J_{2n}^{\rm static}=J_{2n}^{\rm Juno}-J_{2n}^{\rm differential}$ with $J_{2n}^{\rm Juno}$ , that is, the gravitational moments measured by Juno \citep{2020GeoRL..4786572D} and $J_{2n}^{\rm differential}$ , the contribution due to the differential rotation (see \citet{2022A&A...662A..18M}).

In addition, Jupiter's interior is constrained by observations of its atmosphere. The temperature at the 1 bar level was measured first from Voyager radio occultations \citep{Lindal1981} and found to be $T_{\rm 1bar}=165 \pm 5$\,K.
The Galileo entry probe then provided the first and only in situ measurements of Jupiter's atmosphere. The temperature was measured at a single location and found to be $T_{\rm 1bar}=166.1 \pm 0.8$\,K \citep{1998JGR...10322857S}. Nonetheless, \citet{2022PSJ.....3..159G} questioned the representability of this temperature over a wider range of latitudes and longitudes. After reassessing the Voyager radio occultations, these latter authors found that the temperature at 1 bar could reach a value of $T_{\rm 1bar}=170.3 \pm 3.8$\,K. Thus, $T_{\rm 1bar}$ is either fixed or allowed to vary in our models. 

\begin{figure}
   \centering
   \includegraphics[width=\hsize]{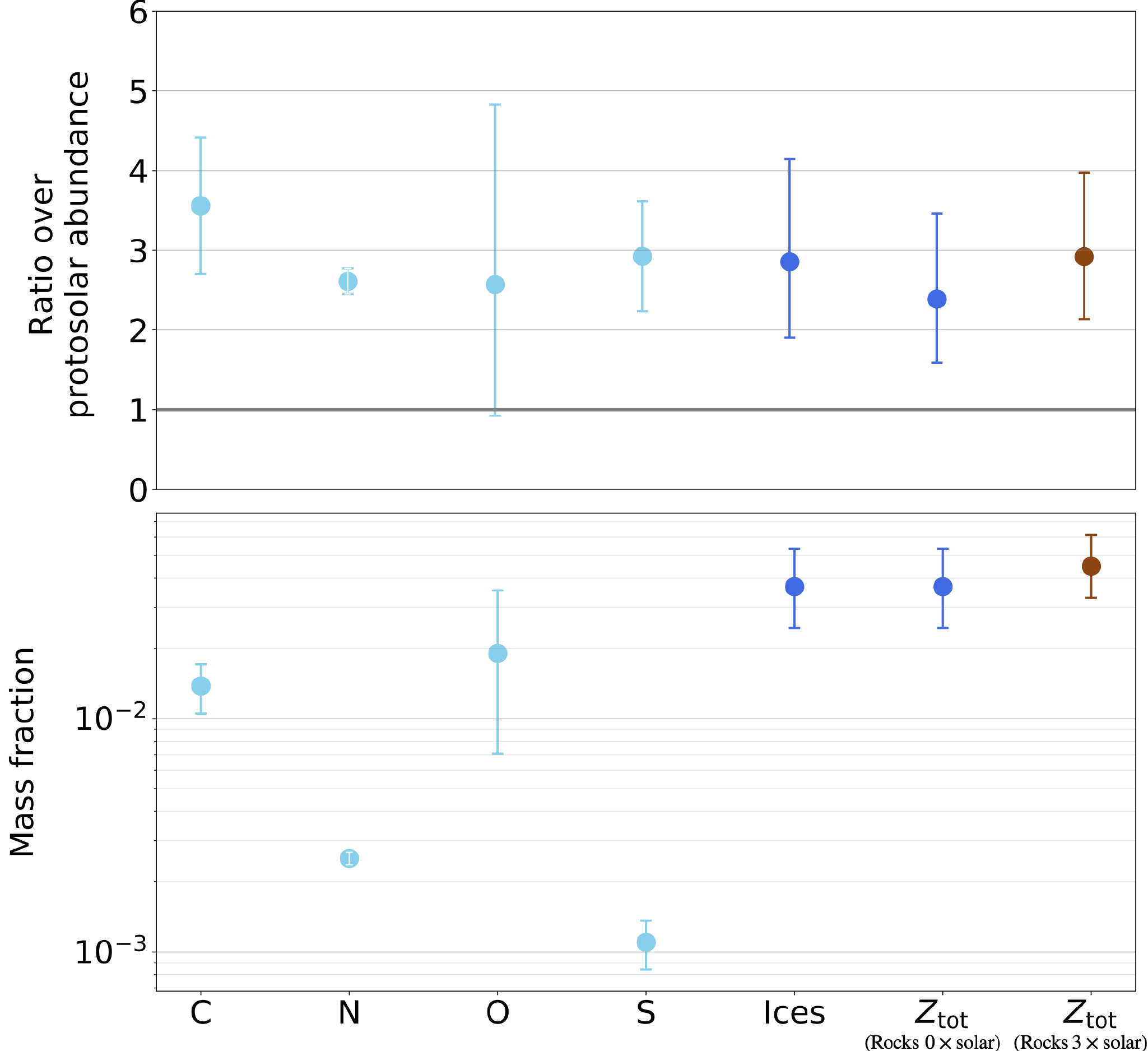}
      \caption{Abundances of heavy elements in the atmosphere of Jupiter. Here, `ices' refers to the metallicity of the  atmosphere considering only ices ($\rm CH_4$, $\rm NH_3$, $\rm H_2O$ and $\rm H_2S$). Two values of $Z_{\rm tot}$ are shown, which correspond to the metallicity of the  atmosphere considering ices (volatiles) and assuming either no rocks (refractories) or an enrichment of three times the protosolar value for rocks. Abundances of methane and hydrogen sulfide are taken from \citet{2004Icar..171..153W}, and ammonia and water abundances are taken from \citet{2020NatAs...4..609L}. Protosolar abundances are taken from \citet{2021A&A...653A.141A}. The protosolar mass fraction of heavy elements is 0.0154. For ices, we calculated the protosolar value considering C, N, O, S, and Ne and included additional elements with the value from \citet{2009LanB...4B..712L}. Concerning rocks, we only considered Mg, Si, and Fe.
              }
         \label{figure:Zatm}
\end{figure}

Galileo also provided measurements of the atmospheric composition. We set the helium abundance in our models through the measurement of $Y_{\rm atm}=Y_{1}/(X_{1}+Y_{1}) = 0.238$ \citep{1998JGR...10322815V} and $Y_{\rm proto}=0.277$ \citep{2010ApJ...719..865S}. The abundances of heavy elements were also measured but the particularly low value of the water abundance from Galileo led to an updated analysis in the equatorial region with the Juno mission \citep{2020NatAs...4..609L}. Figure~\ref{figure:Zatm} sums up the mass fractions of the most abundant heavy elements in Jupiter's atmosphere. We compute the total mass fraction of ices considering only $\rm CH_4$, $\rm NH_3$, $\rm H_2O,$ and $\rm H_2S$ and using a root mean square to estimate its uncertainty. As we do not have measurements of the refractory elements in the atmosphere of Jupiter, we show two values of the total abundance of heavy elements $Z_{\rm tot}$, assuming their abundance to be zero and three times the protosolar value. Depending on the amount of rocks we considered, the total abundance of heavy elements is likely to be between 1.6 ($Z_{\rm tot}=0.024$) and 4.0 ($Z_{\rm tot}=0.061$) times the protosolar abundance. Our models presented in Section~\ref{section:results} use an abundance of 1.3 times the protosolar abundance ($Z=0.02$) which is close to the lower limit of $Z_{\rm tot}$. 

\subsection{Details and calculations of the interior models }
\label{subsec:intmodels_details}

\begin{figure}
   \centering
   \includegraphics[width=\hsize]{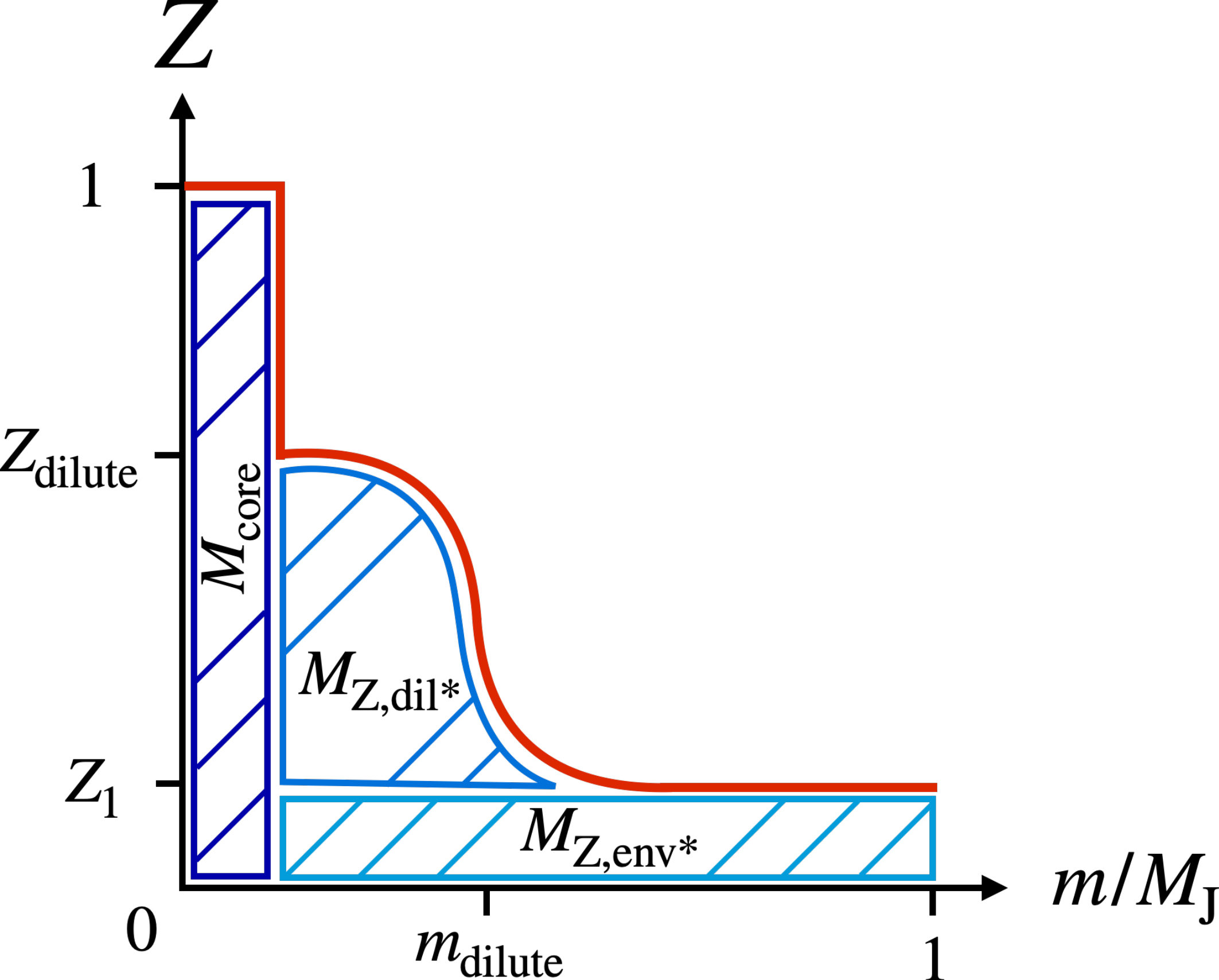}
      \caption{Distribution of heavy elements for a model with a dilute core. The three hashed areas correspond to the mass of the compact core (only made of heavy elements), the mass of heavy elements in the dilute core region excluding the area where $Z<Z_1$, and the mass of heavy elements in the rest of the envelope (where $Z<Z_1$). $Z_1$ is the mass fraction of heavy elements in the outer envelope. $Z_{\rm dilute}$ is the maximum mass fraction in the dilute core. $m_{\rm dilute}$ monitors the extent of the dilute core in terms of mass. 
              }
         \label{figure:schema}
\end{figure}

We keep almost the same parameterisation as in \citet{2022A&A...662A..18M}, with the main difference being that we only consider dilute core models here. Standard three-layer models can be considered as a particular case of dilute core models in which the dilute core is of uniform composition and extends all the way to the helium phase separation region. We note that, as the helium phase separation appears late in the evolution \citep{2020ApJ...889...51M}, it is unlikely that a very strong change in composition can appear in that region. Here we choose not to consider three-layer models and focus on dilute core models.
This latter type of model is composed of a central compact core made exclusively of heavy elements, a dilute core region where the heavy elements are gradually distributed outwards, a metallic hydrogen layer, and an outer molecular hydrogen layer. Figure~\ref{figure:schema} shows the typical distribution of heavy elements in a model with a dilute core. The mass fractions of heavy elements in the molecular hydrogen layer, the metallic hydrogen layer, and the dilute core are respectively monitored with the parameters $Z_{1}$, $Z_{2}$, and $Z_{\rm dilute}$. As a strong change in composition is not expected at the location where helium rain occurs, we fix $Z_{1} = Z_{2}$. We stress that this assumption can affect the gravity harmonics, especially the higher ones, which are more sensitive to the parameters of the outer envelope ($Z_{1}$, $Z_{2}$). The extent of the dilute core is controlled by $m_{\rm dilute}$, which is the normalised mass where the dilute core merges with the metallic hydrogen layer. The mass fraction of heavy elements in the dilute core region is defined by:
\begin{equation}
    Z(m) = Z_1 + \frac{Z_{\rm dilute}-Z_1}{2}
    \left[1- {\rm erf} \left(\frac{m-m_{\rm dilute}}{\delta m_{\rm dil}}\right) \right],
\end{equation}
where $\delta m_{\rm dil}$ is the slope of the gradual change in the mass fraction of heavy elements, and is set to 0.075. The compact core is only made of heavy elements (rocks) and its mass is defined by $M_{\rm core}$. We define two useful quantities, $M_{\rm Z,dil*}$ and $M_{\rm Z,env*}$, in order to assess how predominant the dilute core  is, but also to provide estimates of the amount of heavy elements that needs to be accreted onto the compact core during the formation of the planet.

Finally, we note that we use the code CEPAM \citep{1995A&AS..109..109G} to calculate interior models. We compute the gravitational moments using the theory of figures \citep{1978ppi..book.....Z} at fourth order. The gravitational moments are then calibrated \citep{2018Natur.555..227G} using the accuracy of the concentric MacLaurin spheroid \citep{2012ApJ...756L..15H,2013ApJ...768...43H}. More details can be found in \citet{2022A&A...662A..18M}.

\subsection{The MCMC approach}

We aim to explore a great number of interior models to understand which ones best represent Jupiter's internal structure. We followed the method used in \citet{2022A&A...662A..18M}, which is based on \citet{2012MNRAS.427.1847B}. We took a Bayesian approach based on a Markov chain Monte Carlo (MCMC) code. The equatorial radius as well as the gravitational moments are considered as the data while measurements of the parameters give us a priori information. The MCMC samples models calculated with CEPAM. We computed the posterior density function and therefore the likelihood at various points of the parameter space. We were then able to obtain the joint posterior probability densities. Additional details about the method can be found in the aforementioned papers. Furthermore, in Section~\ref{subsec:priors} we discuss the priors chosen for the parameters of our models and a table with all the parameters is provided in Appendix.

\section{Uncertainty on the H-He equation of state}
  \label{section:eos}
  
\subsection{A diversity of equations of state}

\begin{figure}
   \centering
   \includegraphics[width=\hsize]{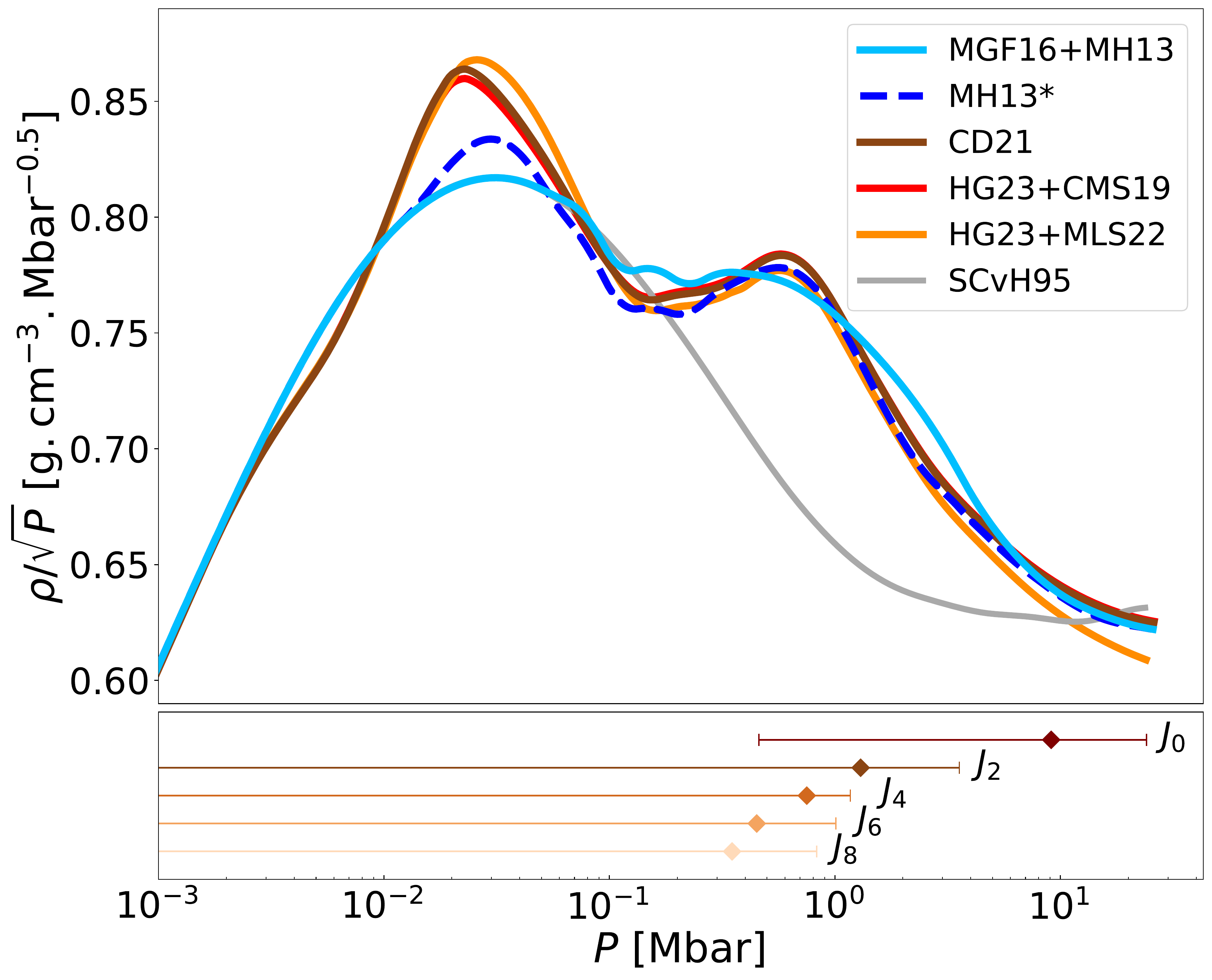}
      \caption{Adiabats obtained from different EOSs and corresponding to a homogeneous model, with no compact core, with $Y=0.245$. Table~\ref{tab:eos} lists the details of the EOSs. Below are shown the contribution functions of the gravitational moments of order 0 to 8 centred at their peak. Their extent corresponds to their full width at half maximum (see \citet{2005AREPS..33..493G} for more details).
              }
         \label{figure:eos_diversity}
\end{figure}

Jupiter being mostly composed of hydrogen and helium, interior models of the planet require the use of an appropriate EOS for these chemical elements. The SCvH95 \citep{1995ApJS...99..713S} EOS was commonly applied for studying Jupiter and extrasolar giant planets \citep{2016ApJ...831...64T}. However, in the last decade, with the improvement of shock Hugoniot data analyses and the development of ab initio simulations, new H-He EOSs have emerged \citep{2013ApJ...774..148M,2019ApJ...872...51C,2022A&A...664A.112M}. 
To compare these EOSs, we derived adiabats for pure hydrogen--helium mixtures ($Y=0.245$) by simply integrating the adiabatic gradient starting from 1\,bar, 166.1\,K (i.e. Jupiter's conditions). For the sake of comparison, heavy elements have not been included but they would only slightly affect the adiabats (see \citet{Helled+2018}).
Table~\ref{tab:eos} details the specifics of the EOSs used to calculate these adiabats. The MGF16+MH13 EOS was derived by \citet{2016A&A...596A.114M} from the ab initio simulation data of \citet{2013ApJ...774..148M}. We built MH13* by fitting an adiabat provided by Burkhard Militzer (private communication), based on the ab initio EOS of \citet{2013ApJ...774..148M}. We used a polynomial function $g$ to fit the residuals between the MGF16+MH13 adiabat and the one provided by BM. We then perturbed the MGF16+MH13 EOS so that $\rho_{\rm MH13*}=(1-g) \rho_{\rm MGF16+MH13}$. The comparison between MGF16+MH13 and MH13* is detailed in Section~\ref{subsec:interp}. More information about the derivation of CD21 can be found in \citet{2021ApJ...913L..21D}. The HG23+CMS19 and HG23+MLS22 EOSs both make use of the tables of the non-linear mixing effects provided in \citet{HowardGuillot2022}. We note that all these EOSs, except SCvH95, account for the interactions between hydrogen and helium particles. Still, MGF16+MH13, CD21, and MH13* include the non-ideal mixing effects but they remain fixed and equal to that calculated by \citet{2013ApJ...774..148M} for a single composition ($Y=0.245$). On the other hand, HG23+CMS19 and HG23+MLS22 include the H-He interactions and remain valid for any composition of the mixture (see \citet{HowardGuillot2022}).
For convenience, as $P \propto \rho^2$  in Jupiter's interior \citep{1975SvA....18..621H}, Figure~\ref{figure:eos_diversity} compares values of $\rho/\sqrt{P}$. The SCvH95 EOS is much less dense than the other, more recent EOSs, ranging from 0.1 to 10\,Mbar. (Quantum Monte Carlo simulations \citep{Mazzola2018PRL} yield even denser hydrogen (between 0.3 and 2.6~Mbar) than found by EOSs obtained from density functional theory.) We find a maximum relative deviation between the different EOSs (except SCvH95) that amounts to 5.5\% (around 0.03~Mbar). The discrepancy between the various EOSs gives us an estimate of the uncertainty on the EOS to be accounted for in Jupiter models. 

\begin{table*}
\centering
\caption{H-He EOSs used in interior models. Concerning the heavy elements, the EOSs used for ices and rocks are respectively those of water and dry sand from \citet{1992_sesame}.}
\begin{tabular}{|l|c|c|p{3.5cm}|p{6cm}|}
\hline
Name of the H-He EOS &  H EOS & He EOS & H-He interactions & Notes/References \\
\hline
\hline
MGF16+MH13  & MGF16-H & SCvH95-He & Included in the H EOS but fixed at $Y=0.245$ & Derived by \text{\cite{2016A&A...596A.114M}} from \text{\cite{2013ApJ...774..148M}} \\ \hline
MH13* & - - & - - & - - & Adjusted from MGF16+MH13 to fit a $Y=0.245$ adiabat from \citet{2022PSJ.....3..185M}
\\ \hline
CD21 & CD21-H & CMS19-He & Included in the H EOS but fixed at $Y=0.245$ & Derived by \text{\cite{2021ApJ...917....4C}} from \text{\cite{2013ApJ...774..148M}} \\ \hline
HG23+CMS19 & CMS19-H & CMS19-He & HG23 & H and He EOSs from \text{\cite{2019ApJ...872...51C}} and non-ideal mixing effects from \text{\citet{HowardGuillot2022}} \\ \hline
HG23+MLS22 & MLS22-H & CMS19-He & HG23 & H EOS from \text{\cite{2022A&A...664A.112M}} and non-ideal mixing effects from \text{\citet{HowardGuillot2022}} \\ \hline
SCvH95  & SCvH95-H & SCvH95-He & None & H and He EOSs from \text{\cite{1995ApJS...99..713S}} \\
\hline
\end{tabular}
\label{tab:eos}
\begin{flushleft}
\end{flushleft}
\end{table*}

\subsection{Interpolation uncertainties}
\label{subsec:interp}

\begin{figure*}
   \resizebox{\hsize}{!}
            {\includegraphics{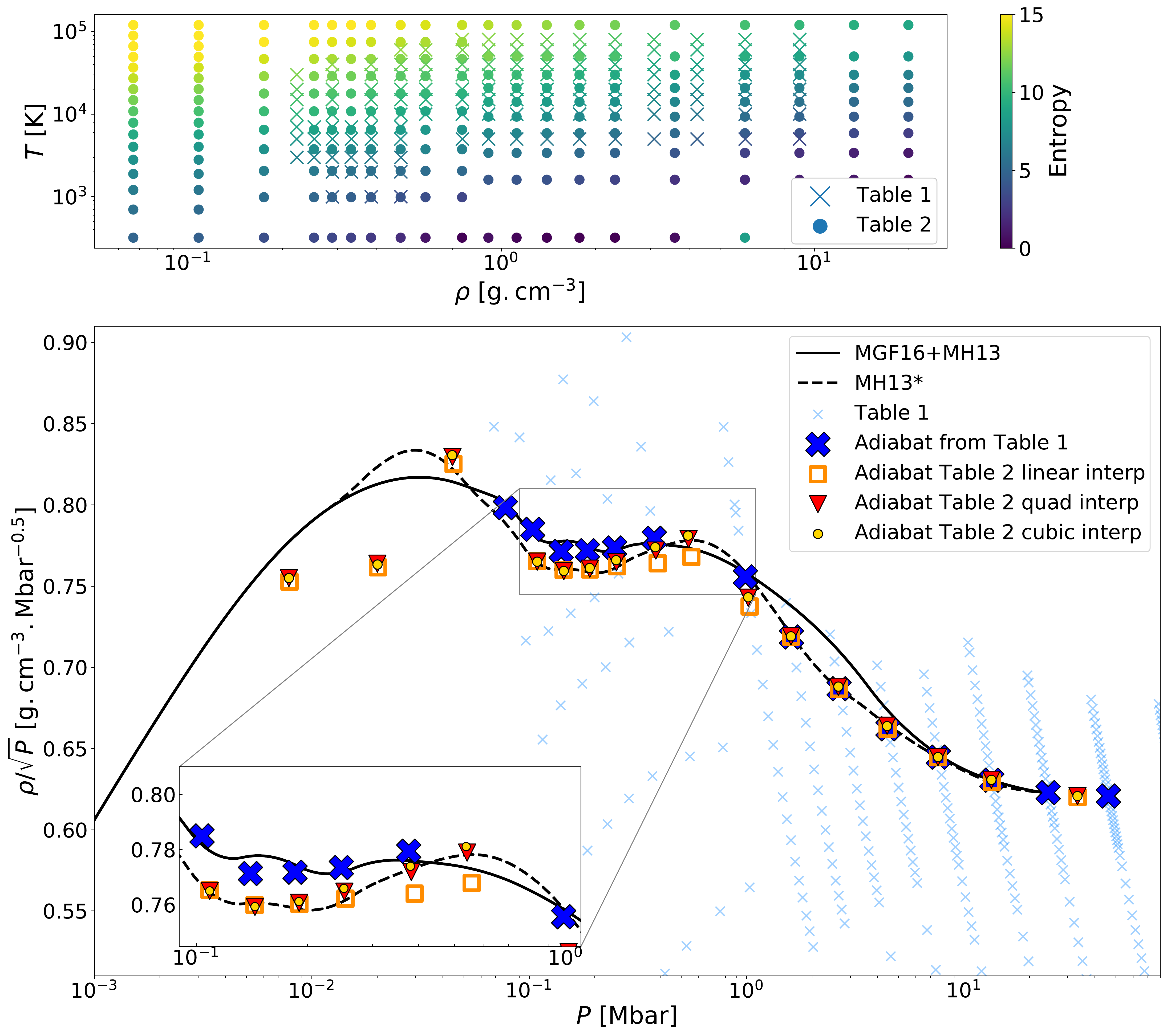}}
      \caption{Comparison between the MH13* and MGF16+MH13 EOSs. \textit{Top panel.} Comparison of Table~1 and Table~2 from \citet{2013ApJ...774..148M}. \textit{Bottom panel.} Comparison of adiabats obtained with the MH13* and MGF16+MH13 EOSs and of points obtained by direct interpolation of the above tables. The $(P,\rho/\sqrt{P})$ points of Table~1 from \citet{2013ApJ...774..148M} are shown as blue crosses. The locations at which the entropy is equal to 7.078061~$k_b \rm el.^{-1}$ (a typical value of entropy for Jupiter) according to the interpolations in the two tables are shown with different symbols, as labelled.
              }
         \label{figure:mh13vsmh13*}
\end{figure*}

The differences between the EOSs seen in Fig.~\ref{figure:eos_diversity} are surprising, because in that parameter space, with the exception of SCvH95, they are all based on the results obtained by the same ab initio calculations from \cite{2013ApJ...774..148M}. In order to understand where these differences come from, we must examine the way the EOS tables are constructed. Two tables are available in \citet{2013ApJ...774..148M}. Table~1 provides the thermodynamic quantities obtained from the density functional molecular dynamics (DFT-MD) calculations. This latter was directly used by \citet{2016A&A...596A.114M} and grafted to the SCvH95 EOS to construct the MGF16+MH13 table. On the other hand, Table~2 from \citet{2013ApJ...774..148M} provides coefficients for a free-energy fit from which one can calculate all thermodynamic quantities. This free-energy fit is used by \citet{2022PSJ.....3..185M} and forms the basis of the MH13 EOS (similar to our MH13* EOS). Both the CD21 EOS \citep{2021ApJ...917....4C} and the nonideal mixing tables of \citet{HowardGuillot2022} use this EOS and have to interpolate the results with the SCvH95 EOS in the molecular regime. To assess the uncertainty due to the choice of table, but also due to the way we interpolate through the table, we derived adiabats from both Table~1 and Table~2 from \citet{2013ApJ...774..148M}. To do so, we used a one-dimensional interpolation to evaluate pressure and temperature at a typical value of entropy for Jupiter (7.078061~$k_b \rm el.^{-1}$), for each density value. This procedure is straightforward for Table~1, but for Table~2 we followed the procedure prescribed by \citet{2013ApJ...774..148M} before deriving the adiabats. We then tried three different types of interpolation (linear, quadratic, and cubic) when calculating the adiabats. Figure~\ref{figure:mh13vsmh13*} shows the different extent in parameter space of the two tables from \citet{2013ApJ...774..148M} and how different choices, in particular on the order of the interpolation, affect the resulting adiabat. Table~2 is slightly extended compared to Table~1 and provides density and entropy for temperatures between 1000 and 80,000~K, and pressures between 0.1 and 300~Mbar. The maximum deviation between adiabats calculated from Table~1 and Table~2 is of the order of 2\%. The order of the interpolation brings a maximum deviation that amounts to 1.3\%. These uncertainties lead to the differences between the MGF16+MH13 and the MH13* EOSs. However, at pressures of lower than 0.1~Mbar, there are also discrepancies between both EOSs that are certainly due to the combination of the DFT-MD calculations of \citet{2013ApJ...774..148M} with the SCvH95 EOS of \citet{1995ApJS...99..713S}. The construction of an EOS is very sensitive to the merging of several tables, particularly around the regions where the tables are connected. This may explain the high values of density around 0.03~Mbar and the slightly lower densities at $P<0.01$~Mbar for CD21, HG23+CMS19, and HG23+MLS22 (see Fig.~\ref{figure:eos_diversity}). Furthermore, we can see that the points of Table~1 ---displayed in Fig.~\ref{figure:mh13vsmh13*}--- are sparse, particularly between 0.1 and 1~Mbar, at densities relevant to the region used to derive a Jupiter adiabat, affecting the accuracy of the interpolation through the table.

\subsection{A thermodynamically consistent modification of the EOS}
\label{subsec:integral_constraint}
\begin{figure}
   \centering
   \includegraphics[width=\hsize]{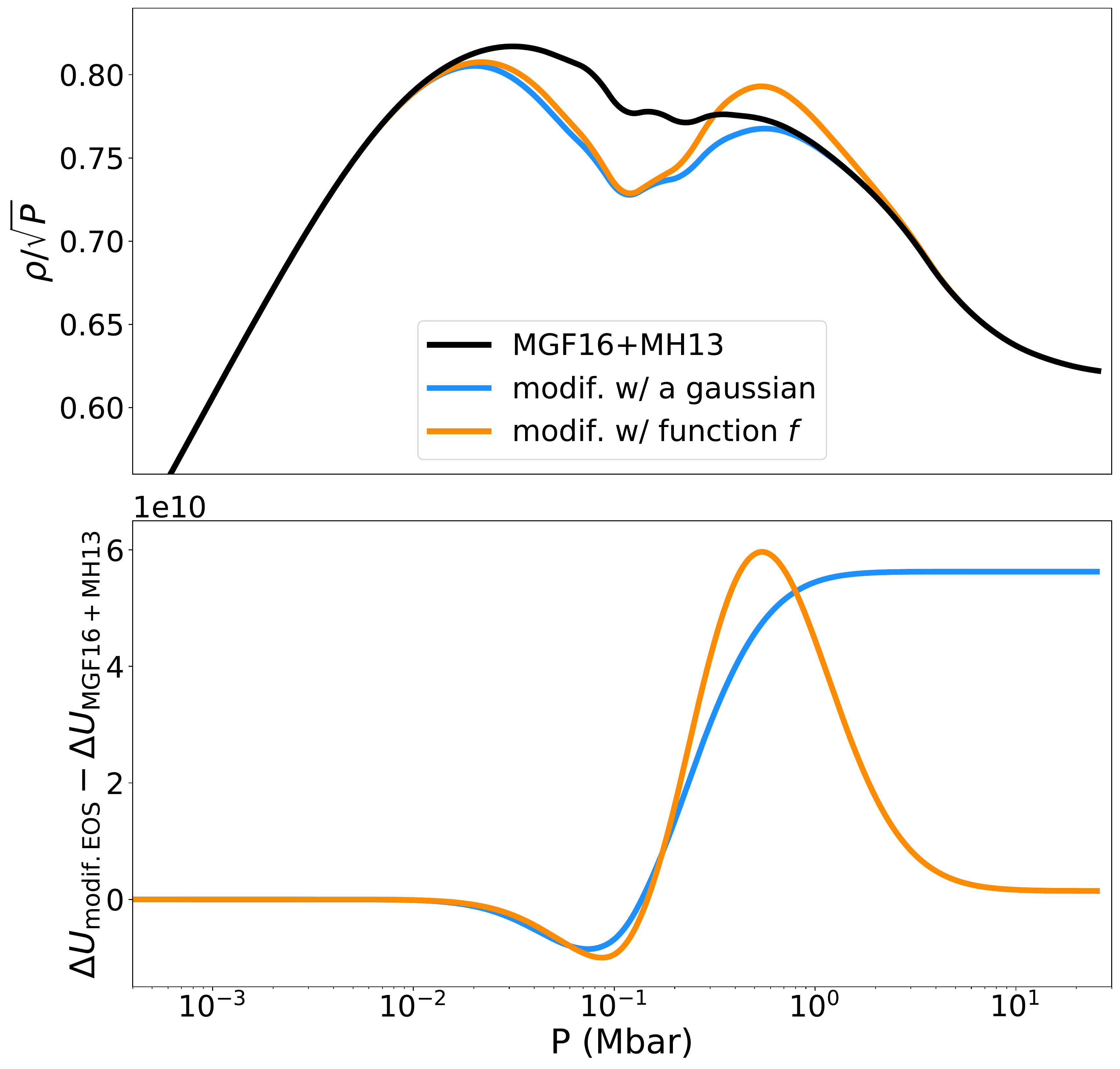}
      \caption{Comparisons of the $\Delta U$ difference for a model modified with only a Gaussian function (blue) and a model modified with the defined function $f$ (see Eq.~\eqref{eq:f}) composed of a Gaussian and an error function (orange). \textit{Top panel.} Comparison of the adiabats ($P_{\rm modif}=10^{11}\,\rm dyn.cm^{-2}$, $\Delta P$=0.5, $d\rho$=-0.06 for the blue curve \& $P_{\rm modif}=10^{11.5}\,\rm dyn.cm^{-2}$, $\Delta P$=0.6, $d\rho$=-0.05). \textit{Bottom panel.} Difference between the $\Delta U$ of the modified models and that of the reference model (MH13).
              }
         \label{figure:deltaU}
\end{figure}

With the uncertainty on the EOSs in hand, we want to account for it in our interior models. To do so, we need a function perturbing an EOS. Initially, we simply used a Gaussian function to perturb the density profile of our models (similarly to \citet{2021PSJ.....2..241N}). Nonetheless, an EOS cannot be perturbed freely. Indeed, any variations of the EOS should satisfy the limits of a thermodynamical potential. The Helmholtz free energy, which is usually where an EOS comes from, is relatively well known at low and high density regimes. Low densities correspond to the regime of an ideal gas and the free energy is known from experimental data and statistical mechanics. High densities correspond to the regime well above the metallisation pressure where hydrogen is fully ionised and the free energy here is known from theory and simulations. Here, we use the internal energy because $P = -\left(\frac{dU}{dV}\right)_S$ can lead to an integral constraint on legitimate density changes of the EOS. If we know the internal energy at low and high regimes for a given entropy, that is, $U(\rho_1,S)$ and $U(\rho_2,S)$, 
respectively, we can obtain an expression of the differences between these two terms, which is to be conserved by perturbations along the adiabat:
\begin{equation}
    \Delta U = U(\rho_2,S) - U(\rho_1,S)
    = \int_{\rho_1}^{\rho_2} \frac{P}{\rho^2}d\rho
    \label{eq:integral_constraint}
.\end{equation}
If $\delta \rho$ corresponds to a slight density modification, then we have
\begin{equation}
    \Delta U_{\rm modif.\ EOS}-\Delta U_{\rm orig.\ EOS} = \int_{\rho_1}^{\rho_2} \frac{P}{(\rho+\delta \rho)^2}d\rho - \int_{\rho_1}^{\rho_2} \frac{P}{\rho^2}d\rho
    \label{eq:deltaU1}
,\end{equation}
Using the approximation $P \propto \rho^2$, we then get
\begin{equation}
    \Delta U_{\rm modif.\ EOS}-\Delta U_{\rm orig.\ EOS} \propto \int_{\rho_1}^{\rho_2} \frac{\delta \rho }{\rho}d\rho = 0
    \label{eq:deltaU2}
.\end{equation}
This difference must be 0 if $\Delta U$ in Eq.~\eqref{eq:integral_constraint} is to be conserved. Equation~\eqref{eq:deltaU2} provides a constraint on how we are allowed to perturb an EOS. Hence, we need to choose an appropriate function to perturb the EOS while verifying this constraint. This function, denoted $f$ here, comes from the equation
\begin{equation}
    \rho_{\rm modif} = \rho + \delta \rho = \rho (1 + f)
.\end{equation}
Therefore, we need to find $f$ so that
\begin{equation}
    \int_{\rho_1}^{\rho_2} f  d\rho = 0
.\end{equation}
Using again $P \propto \rho^2$ and changing variables, we need to choose an $f$ that satisfies
\begin{equation}
    \int_{P_1}^{P_2} f(\rm log_{10}(P))  \sqrt{P}  d\rm log_{10}(P) = 0
.\end{equation}
We naturally define $f$ as
\begin{equation}
    f = K  \frac{d\rho}{\sqrt{P}}  \exp\left(-{\left[\frac{\rm log_{10}(P/P_{\rm modif})}{\Delta P}\right]}^2\right) \rm erf\left(\frac{\rm log_{10}(P/P_{\rm modif)}}{\Delta P}\right)
    \label{eq:f}
.\end{equation}
The function $f$ is composed of a Gaussian and an error function, and includes a division by the square root of the pressure, with $d\rho$ being the amplitude of the density modification. From Eq.~\eqref{eq:deltaU2}, we infer that to satisfy this integral constraint, a density reduction at a certain pressure will imply a density increase at another pressure. To modify an EOS and obtain this trend, we use the product of a Gaussian and an error function. And to properly satisfy the integral constraint from Eq.~\eqref{eq:deltaU2} after changing the integral as a function of the density into an integral as a function of $\rm log_{10}(P)$, we need to multiply by the square root of the pressure.
This function depends on three parameters: $P_{\rm modif}$ corresponds to the pressure (in cgs units) at which the density modification is applied, $\Delta P$ controls the width of the modification; and $d\rho$ is the amplitude of the density change. The constant K is in units of square root of pressure divided by density and is set to $K=1.04 \times 10^6~\sqrt{\rm dyn.cm^{-2}}. \rm g^{-1}.cm^3$.
Figure~\ref{figure:deltaU} shows how the integral constraint from Eq.~\eqref{eq:deltaU2} is satisfied for two different models. One model simply uses a Gaussian function to modify the density profile and  clearly does not respect the integral constraint. On the other hand, the second model, which uses the function $f$ defined above, satisfies this constraint well, because the $\Delta U$ difference falls close to zero at high pressures. More precisely, this value at high pressures is not exactly zero due to the perturbation theory approximation applied to Eq.~\eqref{eq:deltaU1} where $\delta \rho/ \rho$ is assumed small.
But overall, there is a significant difference between how the integral constraint is respected between the two models of Figure~\ref{figure:deltaU}, and the effort of satisfying this constraint must be underlined. We stress that our function $f$ was naturally but arbitrarily chosen; there are certainly other functions that could satisfy Eq.~\eqref{eq:deltaU2}.

\subsection{Priors on the modification of the EOS}
\label{subsec:allowed_modif}

\begin{figure}
   \centering
   \includegraphics[width=\hsize]{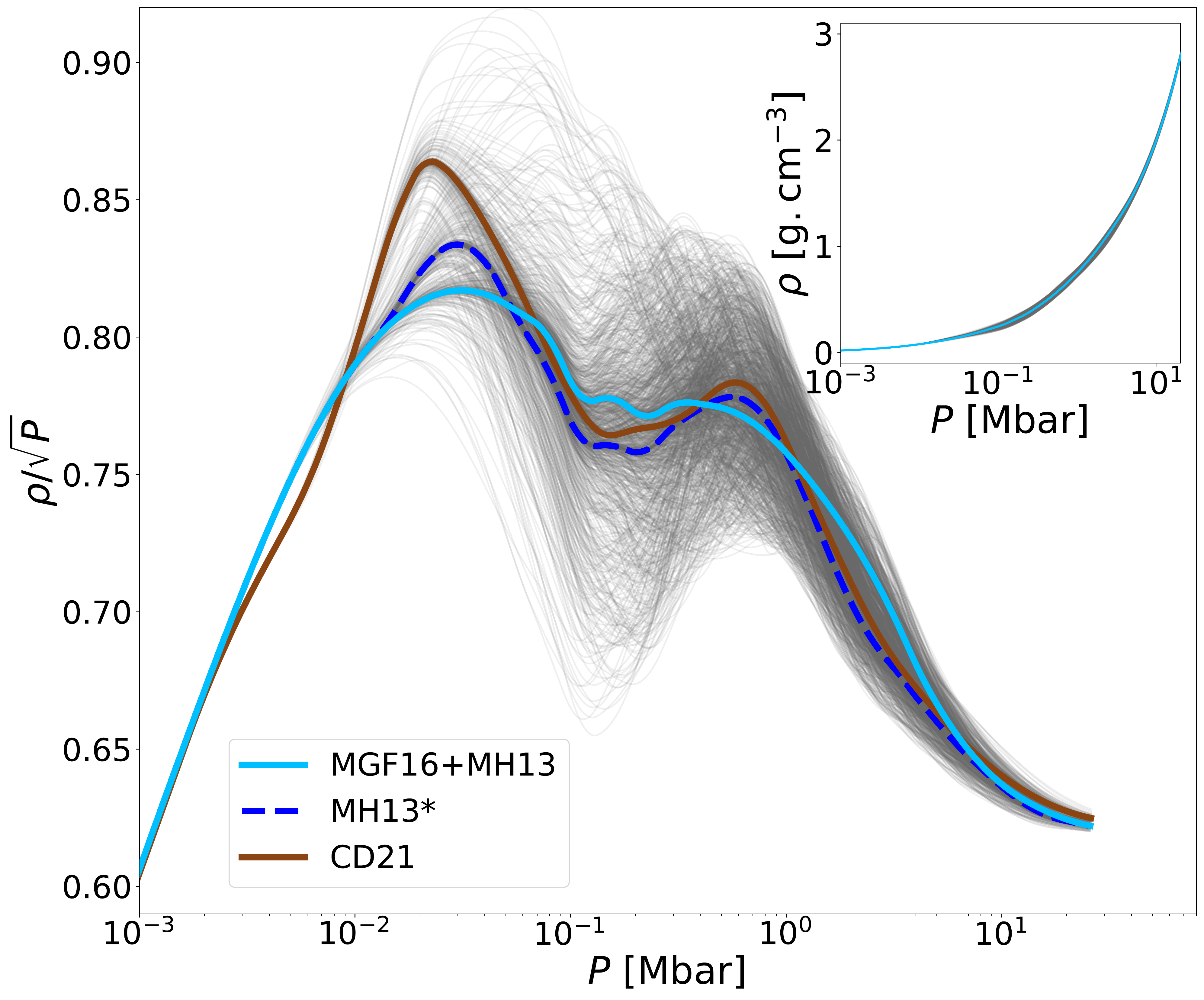}
      \caption{Possible adiabats (gray lines) that can be obtained when modifying the MGF16+MH13, MH13*, CD21, HG23+CMS19, and HG23+MLS22 EOSs with the chosen priors (see Table~\ref{tab:priors}).
              }
         \label{figure:mcmc_priors}
\end{figure}

For MCMC runs in which we allow modifications of the EOS, we use priors on $d\rho$, $P_{\rm modif}$, and $\Delta P$, as defined in Eq.~\eqref{eq:f}. The priors are either Gaussian or uniform (a more detailed discussion can be found in Section~\ref{subsec:runs_modif}) with boundaries set to avoid physically inconsistent modifications. We set $P_{\rm modif}$ between $10^{11.5}$ and $10^{12.5}\,\rm dyn.cm^{-2}$ (which correspond to 0.3 and 3~Mbar), as a preliminary study shows us that a density reduction followed by an increase in the density at higher pressure is preferred over a density increase followed by a decrease. We set $\Delta P$ between 0.2 and 0.8 so that low ($< 10^{-2}$~Mbar) and high ($> 10$~Mbar) pressures are not significantly affected by the density modification. Allowing $P_{\rm modif}$ and $\Delta P$ to pass the chosen boundaries would allow modifications of the EOS that are not consistent with the constraints. The boundaries of $d \rho$ are set to allow a change of amplitude in density that does not exceed 10\%. Using a random sampling, Figure~\ref{figure:mcmc_priors} shows the possible modifications that can be allowed from the original EOSs with the priors and boundaries we chose. When plotting the density according to the pressure, we can see that the differences are small even when we perturb the adiabats. Table~\ref{tab:priors} sums up the priors we chose to run further MCMC simulations.

\section{Results}
  \label{section:results}
  
Constraints on the interior structure and composition of Jupiter are derived as follows. First, as described in the following subsection, in order to avoid results in which the solutions are too far from the observational constraints, we chose to fix some parameters. On this basis, we ran models for the different EOSs and outer envelope metallicity $Z_1$. In the following subsections, we present two types of results: (a) models with the original EOSs, variable $T_{1\,\rm bar}$ and $Z_1=0.02$; and (b) models with modified EOSs, a value of $T_{1\,\rm bar}$ equal to either the Galileo value of 166.1~K or the upper limit from \citet{2022PSJ.....3..159G}, 174.1~K, and values of $Z_1=0.02$, $0.029,$ and $0.035$ (1.3 to $2.3~\times$ protosolar). Our results confirm and extend those of \citet{2022A&A...662A..18M}. When using the same hypotheses, we obtain the same results as \citet{2022PSJ.....3..185M} (see Appendix~\ref{appendix:comp_BM}), and results that are consistent with those presented by \citet{2019ApJ...872..100D}.

\subsection{Choice of priors}
\label{subsec:priors}

The optimisation problem is characterised by two issues: (1) The constraints on Jupiter's gravitational moments are extremely tight, meaning that only a tiny fraction of the parameter space allows for successful models, and (2) the high density of recent H-He EOSs imply that in most cases, fitting only Jupiter's radius and $J_2$ value would require nonphysical negative $Z_1$ or core mass values. In practice, this means that some parameters can be led to values that are significantly offset from their prior.

\begin{figure*}
   \sidecaption
   \includegraphics[width=12cm,clip]{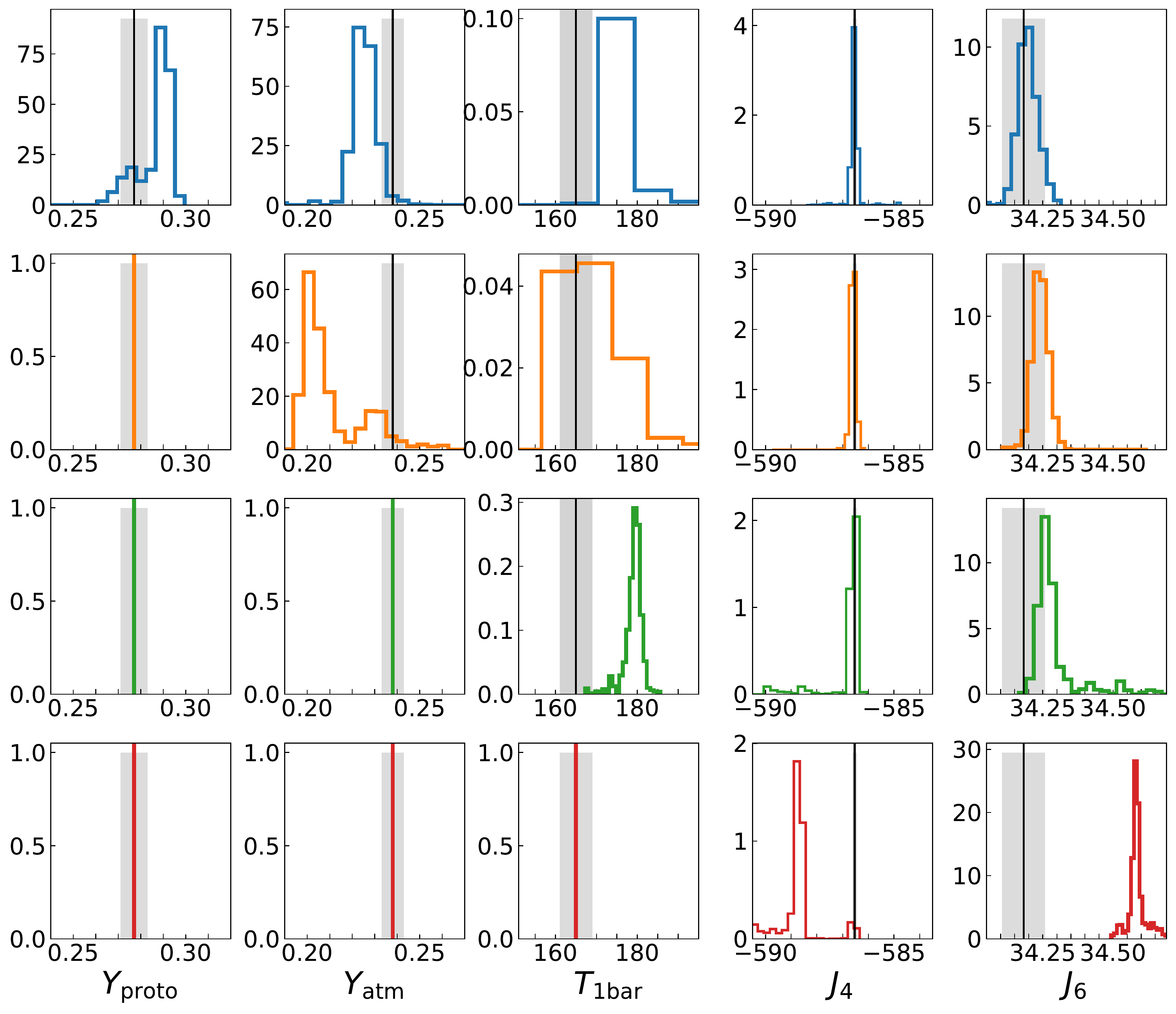}
      \caption{Posterior distributions of $Y_{\rm proto}$, $Y_{\rm atm}$, $T_{\rm 1bar}$, $J_4$, and $J_6$ for four different MCMC runs using the MGF16+MH13 EOS. The blue histograms correspond to a run where the four parameters are free. In the orange run, $Y_{\rm proto}$ is fixed. In the green run, $Y_{\rm proto}$ and $Y_{\rm atm}$ are fixed. In the red run, $Y_{\rm proto}$, $Y_{\rm atm}$, and $T_{\rm 1bar}$ are fixed. When histograms show a black vertical solid line, this indicates that a prior was set and was centred at the value from observations ($Y_{\rm proto}=0.277 \pm 0.006$, $Y_{\rm atm}=0.238 \pm 0.005$, $T_{\rm 1bar}=165 \pm 4$\,K, $J_{4}.10^6=-586.53 \pm 0.0836$ and $J_{6}.10^6=34.18 \pm 0.07682$). The gray areas correspond to the standard deviation of the prior (1 sigma).
              }
         \label{figure:histo}
\end{figure*}

To assess the influence of the priors, we ran four different simulations (using the MGF16+MH13 EOS) with our MCMC code and focus on three parameters: $Y_{\rm proto}$, $Y_{\rm atm}$, $T_{\rm 1bar}$ and two data: $J_4$, $J_6$. In the first run, we set all parameters to vary freely using a Gaussian prior (mean $\mu$, standard deviation $\sigma$) centred on the value from observations. In the second, third, and fourth runs, we respectively fix one, two, and three of the aforementioned parameters. We note that the accuracy on the $J_{2n}$ is clearly higher compared to that on the other parameters. $\sigma/\mu$ is equal to 0.01\% and 0.2\% for $J_4$ and $J_6$ (0.0002\% for $J_2$), respectively, while it is equal to 2\% for the three other parameters. Figure~\ref{figure:histo} shows the posterior distribution of $Y_{\rm proto}$, $Y_{\rm atm}$, $T_{\rm 1bar}$, $J_4$, and $J_6$ for the four different runs. When setting all parameters free, the MCMC code samples models well around the mean values of the $J_{2n}$ measured by Juno (accounting for the influence of differentially rotating winds; see \citet{2022A&A...662A..18M}). However, $Y_{\rm proto}$, $Y_{\rm atm}$, and $T_{\rm 1bar}$ are at 2$\sigma$ or 3$\sigma$ from the mean value of their respective prior. When fixing $Y_{\rm proto}$, the sampled values of $J_6$ are now at 1$\sigma$ from the observed mean value and sampled values of $Y_{\rm atm}$ are at 7-8$\sigma$ from Galileo's measurement. When fixing $Y_{\rm proto}$ and $Y_{\rm atm}$, the $J_6$ fit is slightly poorer than the previous run. However, $T_{\rm 1bar}$ has now a 4$\sigma$ difference from the mean value of the prior. Finally, when fixing $Y_{\rm proto}$, $Y_{\rm atm}$, and $T_{\rm 1bar}$, $J_4$ differs by more than 20$\sigma$ from Juno's measurement and $J_6$ differs by 5$\sigma$.

{This shows that to satisfy the observational constraints on parameters such as $Y_{\rm proto}$ and $Y_{\rm atm}$, we need to fix them instead of using a Gaussian prior, because the uncertainties on the gravitational moments dominate. Otherwise, these parameters will be further than 1$\sigma$ from the observed value. As the observational constraint is looser on $T_{\rm 1bar}$ because of the question of latitudinal dependency ---which raises the possibility that Jupiter's deep temperature may be higher than the Galileo probe reference value (see Section~\ref{subsec:observable_data})---, only $T_{\rm 1bar}$ will be set as a free parameter in some of the runs presented in this paper.}

\subsection{Runs with a modified EOS}
\label{subsec:runs_modif}

\begin{figure}
   \centering
   \includegraphics[width=\hsize]{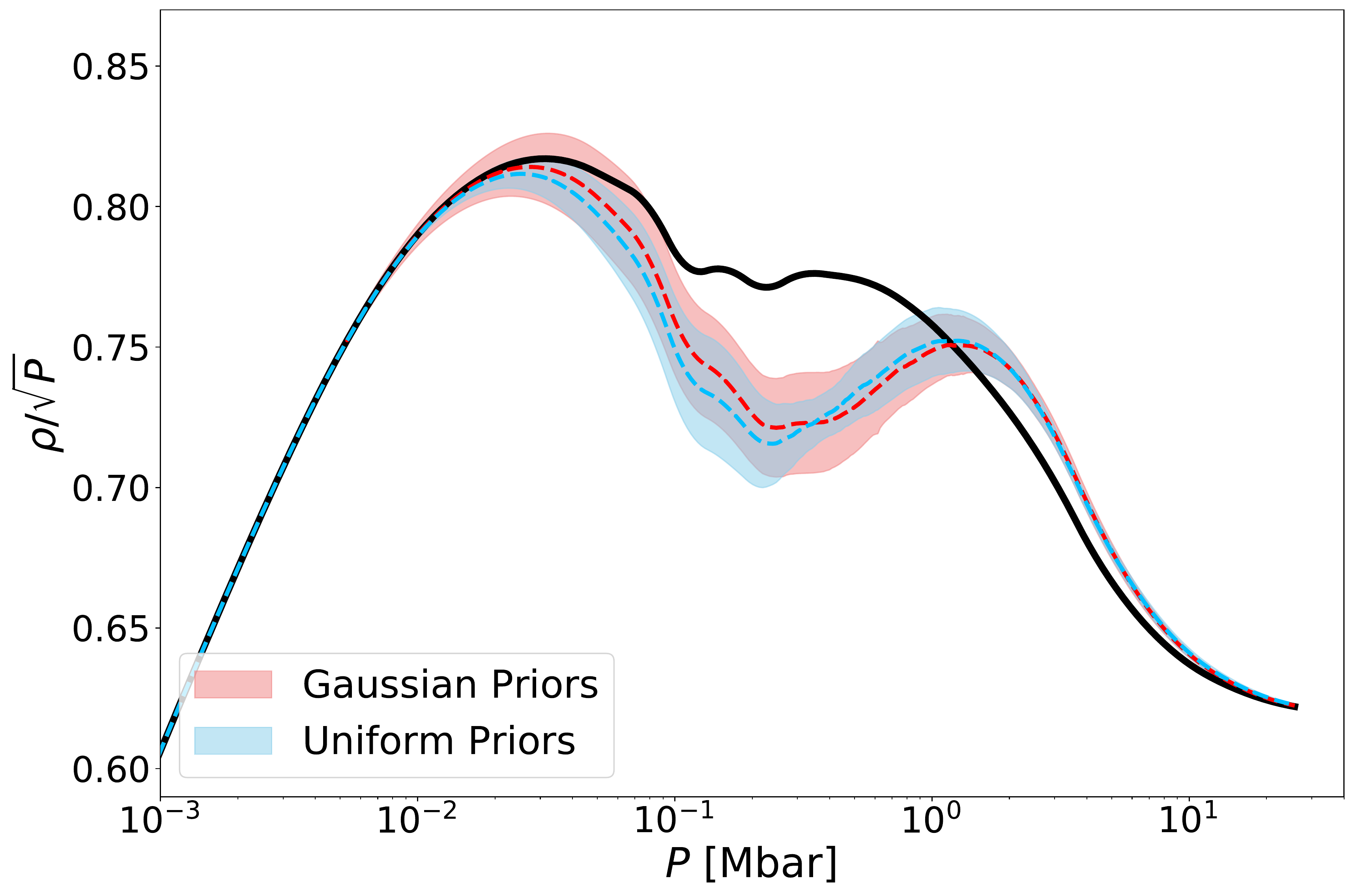}
      \caption{Adiabats obtained for models with a modification of the EOS. The black solid line corresponds to the original MGF16+MH13 EOS. Red shows results obtained with Gaussian priors on $\Delta P$ ($\mu = 0.5$ and $\sigma=0.02$) and $d\rho$ ($\mu = 0.$ and $\sigma=0.01$). Blue shows results obtained with uniform priors. The prior on $P_{\rm modif}$ remains uniform in both cases. The dashed lines correspond to the adiabat obtained with the mean values of $P_{\rm modif}$, $\Delta P$, and $d\rho$ (see Section~\ref{subsec:integral_constraint}) of a subsample of 100 models randomly drawn from the MCMC output. We compute the standard deviation ($\sigma$) of the 100 adiabats and the envelopes show the adiabats of the 1~$\sigma$ spread from the mean modified adiabat (dashed line). Here, $T_{\rm 1bar}$ is fixed at 166.1~K.
              }
         \label{fig:unif_vs_gauss}
\end{figure}

For MCMC runs in which we allow for modifications of the EOS, we first chose Gaussian priors on $P_{\rm modif}$, $\Delta P$, and $d\rho$ (see Sections~\ref{subsec:integral_constraint} and~\ref{subsec:allowed_modif}). Our goal was to penalise models with a substantial change of the EOS and favour models with only a slight modification of the EOS; as for the other parameters discussed in the previous section, this led to large deviations of the EOS. Figure~\ref{fig:unif_vs_gauss} compares the adiabats of models including EOS modifications, with a uniform prior on $P_{\rm modif}$ and either uniform or Gaussian priors on $\Delta P$ and $d\rho$. For the Gaussian priors, the parameters for $\Delta P$ were $\mu = 0.5$ and $\sigma=0.02$, and the parameters for $d\rho$ were $\mu = 0.$ and $\sigma=0.01$. The difference between $\rho/\sqrt(P)-P$ profiles obtained after modification of the EOS is subtle between runs with Gaussian and uniform priors. This slight difference between the modified adiabats leads to a better fit of the data (equatorial radius, gravitational moments) for models obtained with uniform priors. When using Gaussian priors, most of the models are at 2$\sigma$ or 3$\sigma$ from the observed equatorial radius and $J_4$ and are at 4$\sigma$ to 5$\sigma$ from the mean value assumed for the prior on $P_{\rm He}$ (see Fig.~\ref{appendix:cornerplot_unifvsgauss}). As the modified adiabats are very similar and the agreement with the observational constraints is slightly better, we present results obtained with uniform priors.

\begin{figure}
   \centering
   \includegraphics[width=\hsize]{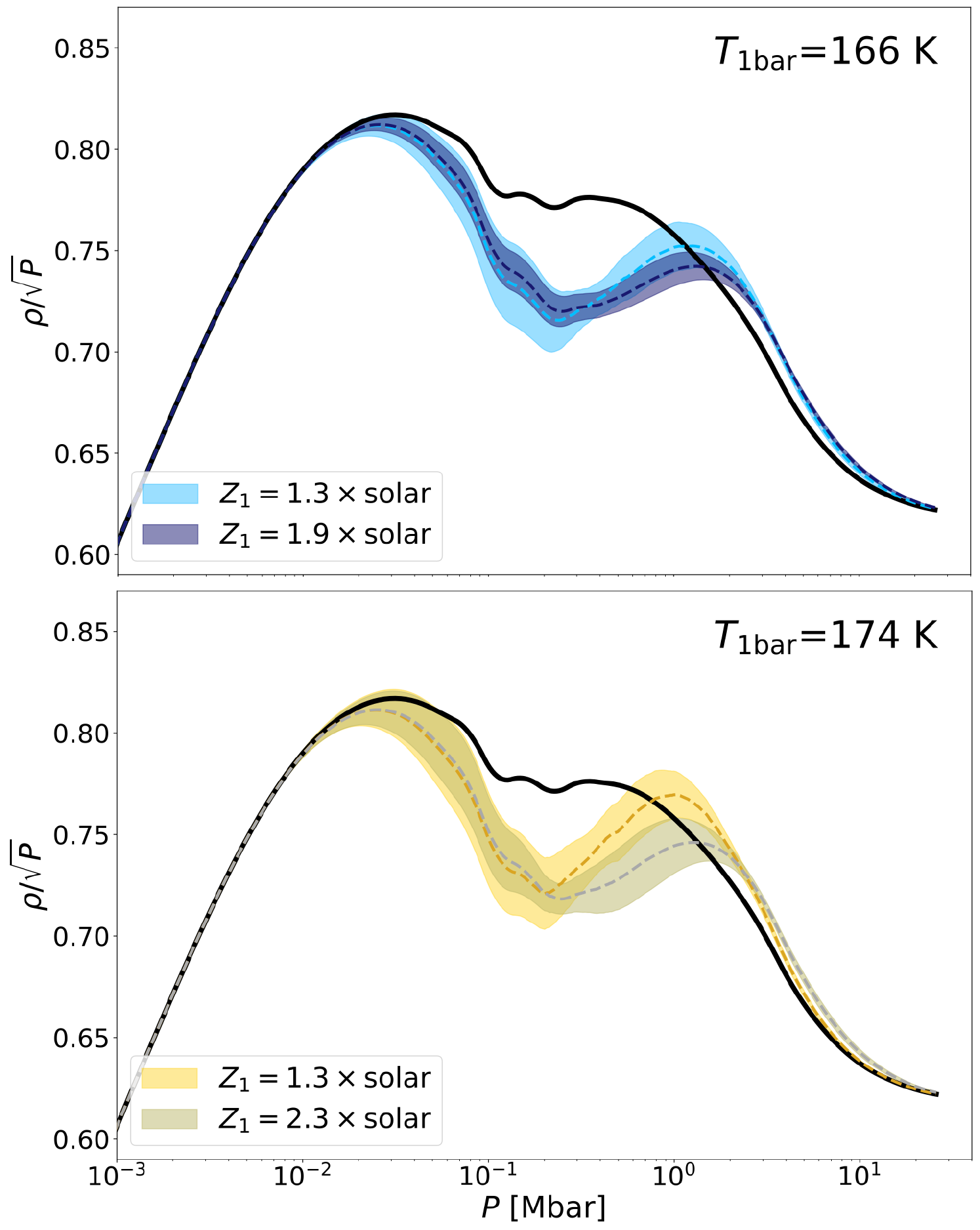}
      \caption{Adiabats obtained for models with modified EOS. \textit{Top Panel.} $T_{\rm 1bar}$ is fixed at 166.1~K. The light blue area shows results for $Z_{1}=0.02$ ($1.3~\times$ the protosolar value) while dark blue shows results for $Z_{1}=0.0286$ ($1.9~\times$ the protosolar value).
      \textit{Bottom Panel.} $T_{\rm 1bar}$ is fixed at 174.1~K. The yellow area shows results for $Z_{1}=0.02$ ($1.3~\times$ the protosolar value) while gray shows results for $Z_{1}=0.035$ ($2.3~\times$ the protosolar value). The black solid line corresponds to the original MGF16+MH13 EOS.
      Other details of the figure can be found in the caption of Fig.~\ref{fig:unif_vs_gauss}.
              }
         \label{fig:modif_adiabats_T166_T174}
\end{figure}

As previously mentioned, the modification of the initial MGF16+MH13 EOS is relatively substantial, with a change in density that can reach up to $\sim 11\%$ in amplitude (see~Fig.\ref{fig:unif_vs_gauss}). Figure~\ref{fig:modif_adiabats_T166_T174} shows the modifications of the EOS for models (using uniform priors) at respectively $T_{\rm 1bar}=166.1$~K and 174.1~K. At higher $Z_1$, the modifications of the EOS occur at higher pressures but in any case, the amplitude remains significant (between 6 and 11\%). In addition, these changes to the EOS are likely to be incompatible with Hugoniot data \citep{2017PhRvL.118c5501K}. We therefore provide the results with modified EOS as a way to test the robustness of the solutions, but we generally focus on results using the original EOSs.

\subsection{Surface temperature $T_{\rm 1bar}$ and helium transition pressure $P_{\rm He}$}
\label{subsec:T1bar_PHe}

\begin{figure}
   \centering
   \includegraphics[width=\hsize]{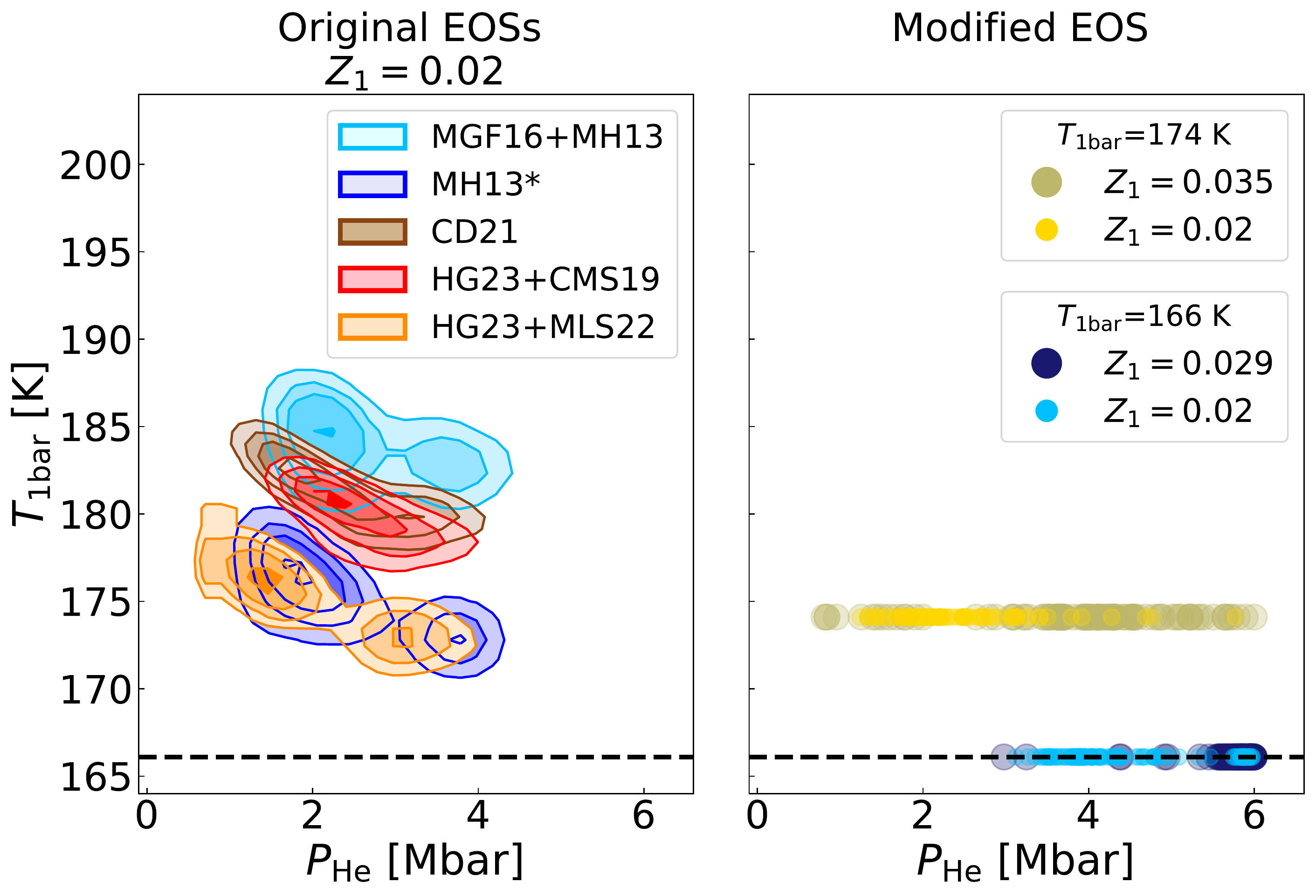}
      \caption{Temperature at 1 bar vs. helium transition pressures for two types of models. \textit{Left panel.} Models using original EOSs. $Z_1=0.02$. \textit{Right panel.} Models with a modification of the EOS. The initial EOS that has been modified is MH13. We present two subsets of models: with $T_{\rm 1bar}=166.1$~K and $T_{\rm 1bar}=174.1$~K. The black dotted line corresponds to the 1 bar temperature measured by the Galileo probe.
              }
         \label{figure:T1bar_PHe}
\end{figure}

As mentioned in the preamble of Section~\ref{section:results}, we present two sets of models: with original EOSs and with a modification of the EOS. Here, we focus on the 1 bar temperature, which prescribes the entropy inside Jupiter, and on the pressure where helium rain occurs \citep{1977ApJS...35..239S}. The latter sets the limit between the molecular hydrogen (helium-poor) and the metallic hydrogen (helium-rich) layers. Figure~\ref{figure:T1bar_PHe} shows the values of these two parameters for the two types of interior models sampled by our MCMC code. Interior models with the original EOSs (and $Z_{\rm atm}=0.02$) all yield a 1 bar temperature, which is higher than the value measured by Galileo, which ranges from 171~ to 188~K. In particular, with the MGF16+MH13 EOS, we obtain a $T_{\rm 1bar}$ of between 180 and 188~K, while we were obtaining a $T_{\rm 1bar}$ of between 175 and 183~K for $Z_{\rm atm}=0.0153$ (protosolar value) in \citet{2022A&A...662A..18M}. Therefore, a Z abundance of $1.3 \times \rm protosolar$ instead of $1 \times \rm protosolar$ in the outer envelope leads to a 5~K increase in $T_{\rm 1bar}$ to fit the $J_{2n}$ (see Section~\ref{subsec:Js}). Only models using MH13* or HG23+MLS22 seem to be in line with the upper end of the temperature provided by \citet{2022PSJ.....3..159G}. Overall, all models present a high 1 bar temperature, which could correspond to a deep entropy in line with a hotter interior due to a potential superadiabaticity \citep{Guillot1995,2012A&A...540A..20L}. The models using original EOSs exhibit a helium transition pressure of between 0.8 and 4.5~Mbar, which is in agreement with the values obtained by simulations and experiments (see \citet{PhysRevB.84.235109,PhysRevB.87.174105,PhysRevLett.120.115703,2021Natur.593..517B}). For all EOSs, we can distinguish two ensembles of solutions: one with high $P_{\rm He}$ corresponding to models with a compact core of a few earth masses (1-6~$M_{\oplus}$) and a highly extended dilute core ($m_{\rm dilute}$ between 0.4 and 0.6) and a second one with lower $P_{\rm He}$ corresponding to models with almost no compact core and a less extended dilute core ($m_{\rm dilute}$ between 0.15 and 0.45) (more details in Section~\ref{subsec:zdil_mdil}). The difference in $T_{\rm 1bar}$ between the two ensembles of solutions is only of 2-3~K. Therefore,  accurate constraint of the  pressure at which helium rain occurs could help to characterise the dilute core and to determine the atmospheric entropy of Jupiter.

Concerning models with a modification of the EOS, we calculate two subsets of models where we fix $T_{\rm 1bar}$ at 166.1~ or 174.1~K. For each temperature, we present results for two values of the abundance of heavy elements in the outer envelope $Z_1$. At $T_{\rm 1bar}=166.1$~K, models with $Z_1=0.02$ have $P_{\rm He}$ values concentrated between 3 and 4.5~Mbar. With $Z_1=0.029$, models tend to high transition pressures, around 6~Mbar, and are far from fitting the observed equatorial radius and gravitational moments (see Section~\ref{subsec:Js}). At $T_{\rm 1bar}=174.1$~K, we obtain values of $P_{\rm He}$ of between 1.5 and 3.5~Mbar for $Z_1=0.02$ and of between 3 and 4.5~Mbar for $Z_1=0.035$, which are both close to what is expected.

\subsection{Equatorial radius $R_{\rm eq}$ and gravitational moments $J_{2n}$}
\label{subsec:Js}

\begin{figure}
   \centering
   \includegraphics[width=\hsize]{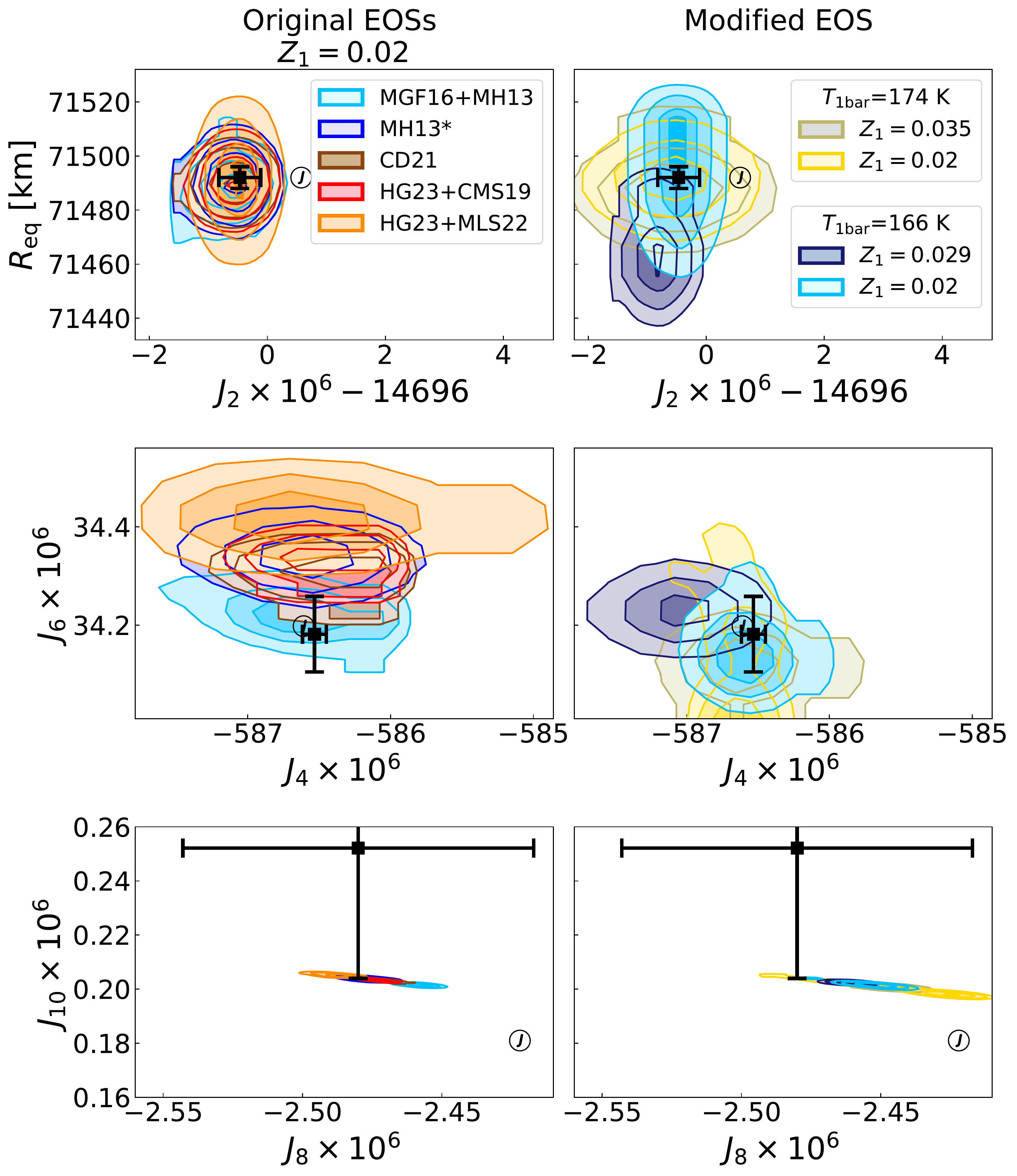}
      \caption{Equatorial radius and gravitational moments for two types of models. \textit{Left panels.} Models using original EOSs. $Z_1=0.02$. \textit{Right panels.} Models with a modification of the EOS. The inital EOS that has been modified is MH13. We present two subsets of models: with $T_{\rm 1bar}=166.1$~K and $T_{\rm 1bar}=174.1$~K. The circles with a $J$ correspond to the measurements of the gravitational moments by Juno \citep{2020GeoRL..4786572D}. The black error bars correspond to the gravitational moments corrected by differential rotation (see \ref{subsec:observable_data}).
              }
         \label{figure:Js}
\end{figure}

Here, we examine the fit of our models to the gravitational moments measured by Juno and accounting for differential rotation. Figure ~\ref{figure:Js} shows the equatorial radius and the gravitational moments obtained with our interior models. All models with original EOSs can reproduce the equatorial radius and all the gravitational moments except $J_6$. We find solutions  for MGF16+MH13  that can match $J_6$ corrected by differential rotation. For the four other EOSs, the sampled values of $J_6$ are in the 2 - 3~$\sigma$ range. We can see a correlation between $T_{\rm 1bar}$ and $J_6$: models using an EOS that yields higher $T_{\rm 1bar}$ present lower values of $J_6$. 
{With MGF16+MH13, $J_6 \times 10^6$ is between 34.1 and 34.3. With MH13*, CD21 and HG23+CMS19, $J_6 \times 10^6$ is between 34.2 and 34.4. With HG23+MLS22, $J_6 \times 10^6$ is between 34.3 and 34.5. \citet{2022PSJ.....3..185M} found a value of $J_6 \times 10^6$ of 34.47 for $T_{\rm 1bar}=166.1$~K using their EOS from \citet{2013ApJ...774..148M}.}

Concerning models allowing for a modification of the EOS, at $T_{\rm 1bar}=166.1$~K, we manage to find models matching $R_{\rm eq}$ and all $J_{2n}$ when $Z_1=0.02$. However, at $Z_1=0.029$, we can no longer fit $R_{\rm eq}$ or $J_{2n}$. These models have an equatorial radius that is several sigma below the observational constraint but we still retain them to test the robustness of our results. We then set $T_{\rm 1bar}$ to 174.1~K and find models reproducing $R_{\rm eq}$ and all $J_{2n}$, even for $Z_1=0.035$ ($2.3~\times$ protosolar).

\subsection{Heavy-element distribution}
\label{subsec:heavies_distrib}
\begin{figure}
   \centering
   \includegraphics[width=\hsize]{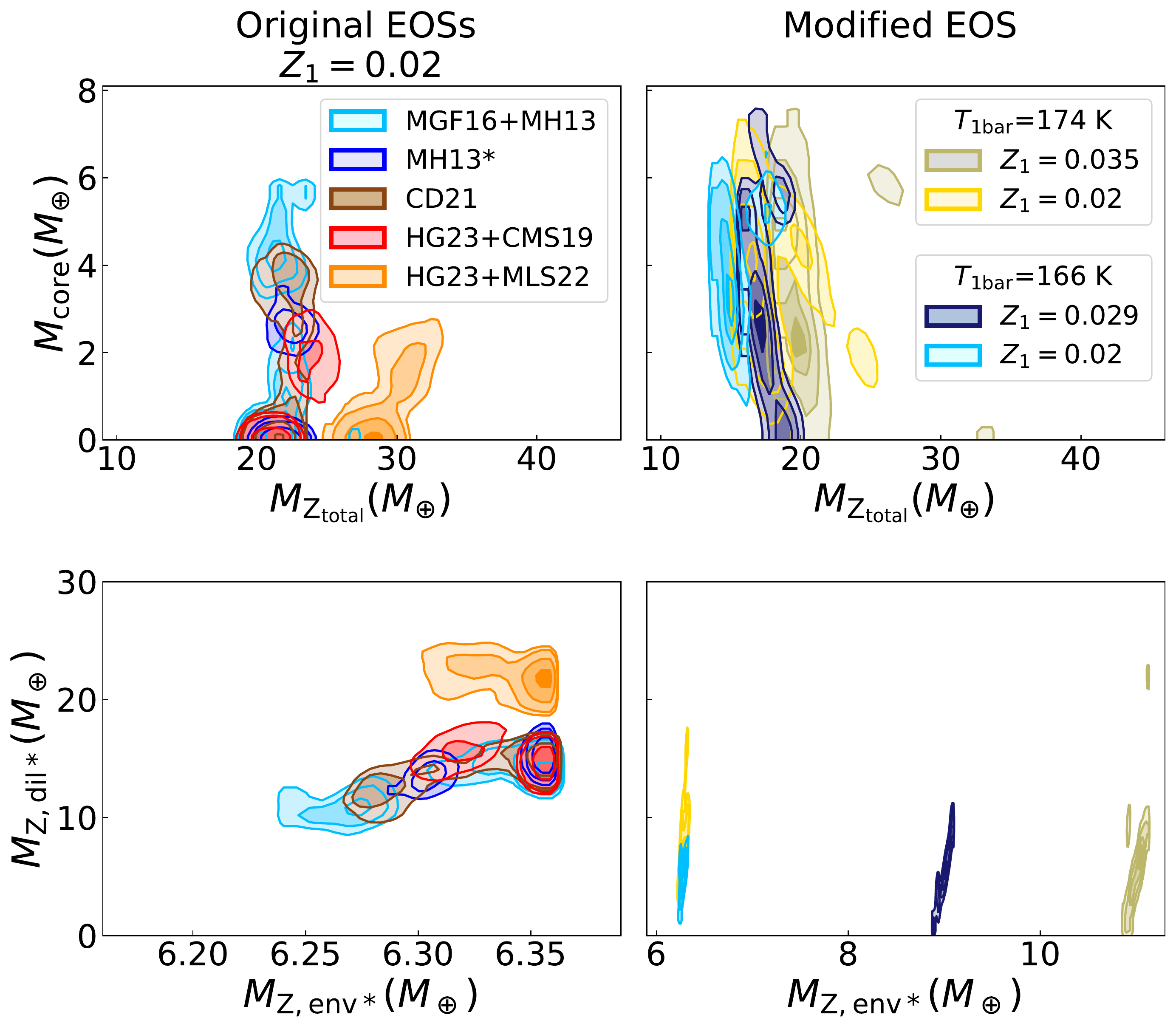}
      \caption{Masses of heavy elements in our interior models. \textit{Top panels.} Mass of the compact core vs. total mass of heavy elements in Jupiter. \textit{Bottom panels.} $M_{\rm Z,dil*}$ vs. $M_{\rm Z,env*}$ (see Section~\ref{subsec:intmodels_details}).
              }
         \label{figure:MZ}
\end{figure}

We now compare the distribution of heavy elements in our models. Figure~\ref{figure:MZ} shows the heavy elements masses defined in Section~\ref{subsec:intmodels_details}. Models with original EOSs have a total mass of heavy elements of between 18 and 33~$M_{\oplus}$ and a compact core of less than 6~$M_{\oplus}$. These results are in line with those obtained by \citet{2022A&A...662A..18M}), showing that increasing the Z abundance in the outer envelope from $1 \times \rm protosolar$ to $1.3 \times \rm protosolar$ does not lead to a drastic change in the distribution of heavy elements. $M_{\rm Z,dil*}$ is between 10 and 25~$M_{\oplus}$, which is larger than $M_{\rm Z,env*}$ by up to a factor of 4. Hence, models with no modification of the EOS have most of their heavy elements in the dilute core region rather than in the rest of the envelope. 

Allowing for modifications of the EOS generally leads to a lower total mass of heavy elements, mostly between 12 and 20~$M_{\oplus}$. This is due to modifications of the EOS that make the adiabats (hence the H-He mixture) denser at depth. The mass of the compact core does not exceed 8~$M_{\oplus}$ for these models. $M_{\rm Z,dil*}$ is similar in all of our four cases: models are concentrated around a region where $M_{\rm Z,dil*} \sim 5-7~M_{\oplus}$. However, $M_{\rm Z,env*}$ clearly depends on the value of $Z_1$. For $Z_1=0.02$, $M_{\rm Z,env*}$ is around 6~$M_{\oplus}$, for $Z_1=0.029$, $M_{\rm Z,env*}$ is around 9~$M_{\oplus}$, and for $Z_1=0.035$, $M_{\rm Z,env*}$ is around 11~$M_{\oplus}$. Therefore, our models with a modified EOS do not lead to a dilute core that is predominant in heavy elements compared to the rest of the envelope, contrary to what we find for models with original EOSs.

We stress that the total masses of heavy elements inferred here are lower limits: The presence of compositional gradients implies that parts of the interior may be super-adiabatic because it is Ledoux-stable, double-diffusive \citep[see][]{2012A&A...540A..20L}, or stable to moist convection \citep{Guillot1995, Leconte+2017}, meaning that the interior could be warmer and thus retain more heavy elements than calculated here. For example, \citet{2022PSJ.....3..185M} estimated that a doubling of the central temperature of Jupiter would increase the mass of heavy elements from 25 to 42~$M_{\oplus}$.

\subsection{Dilute core characteristics}
\label{subsec:zdil_mdil}

A key question is the extent of this dilute core, which connects interior models of Jupiter and formation and evolution models. Recent interior models from \citet{2022PSJ.....3..185M} and \citet{2019ApJ...872..100D} suggest considerably extended dilute cores that respectively reach 63\% and 65\%-75\% of Jupiter's radius. These values correspond to $\sim 50\%$ and $\sim$~60\%-75\% of Jupiter's mass, respectively, which can be compared to the value of our $m_{\rm dilute}$ parameter. We note that the comparison is not exact because it is affected by the functional form chosen for the dilute core (see Section~\ref{subsec:intmodels_details}), but the effect is minor: In our case, the added heavy element mass fraction in the dilute core drops from 50\% of its maximal value at $m=m_{\rm dilute}$ to only 8\% at $m_{\rm dilute}+\delta m_{\rm dil}$ with $\delta m_{\rm dil}=0.075$.

\begin{figure}
   \centering
   \includegraphics[width=\hsize]{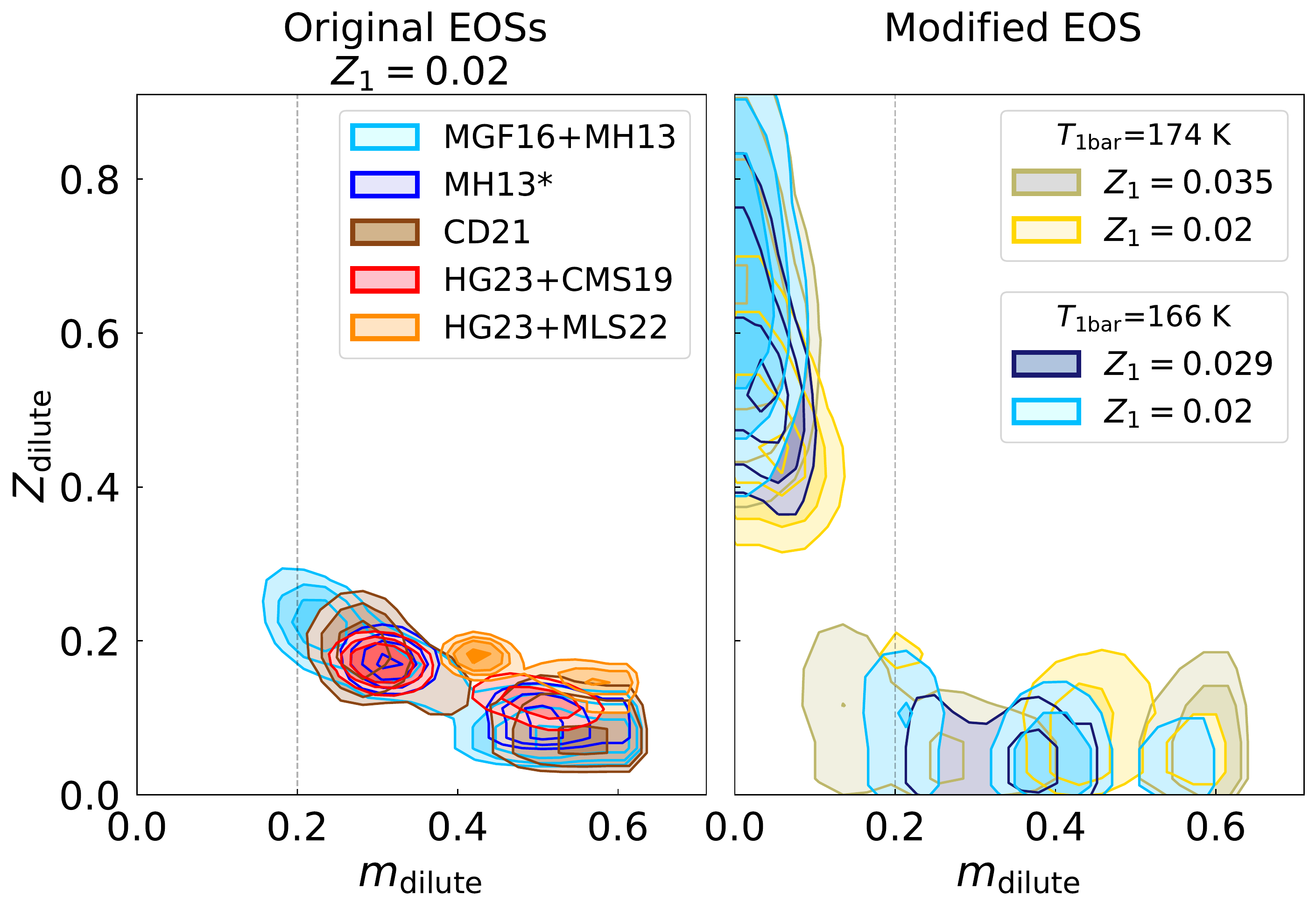}
      \caption{$Z_{\rm dilute}$ vs. $m_{\rm dilute}$ for two types of models. $Z_{\rm dilute}$ is the maximum mass fraction of heavy elements in the dilute core region while $m_{\rm dilute}$ controls the extent of the dilute core in terms of mass (see Section~\ref{subsec:intmodels_details}). \textit{Left panel.} Models using original EOSs. $Z_1=0.02$. \textit{Right panel.} Models with a modification of the EOS. The initial EOS that has been modified is MH13. We present two subsets of models: with $T_{\rm 1bar}=166.1$~K and $T_{\rm 1bar}=174.1$~K. The thin gray dotted line corresponds to $m_{\rm dilute}=0.2,$ which corresponds approximately to the extent of the dilute core of 20\% of Jupiter's mass predicted by formation models from \citet{2020A&A...638A.121M}.
              }
         \label{figure:Zdil_Mdil}
\end{figure}

The preferred model from \citet{2022PSJ.....3..185M} would have a value of $m_{\rm dilute} \sim 0.36$  in our parameterisation. Figure~\ref{figure:Zdil_Mdil} shows the values of $m_{\rm dilute}$ found for our models. Using original EOSs, we obtain $m_{\rm dilute}$ of between 0.15 and 0.6 (the dilute core extends from $\sim$~15\% to $\sim$~60\% of Jupiter's total mass). We confirm that we find models with very extended dilute cores as in \citet{2019ApJ...872..100D} and \citet{2022PSJ.....3..185M}. But we are also finding models with relatively narrow dilute cores (down to $\sim$~15\% of Jupiter's mass). These solutions with relatively small, dilute cores are in better agreement with the mixing and evolution calculations of \citet{2020A&A...638A.121M}, which yield a dilute core that does not exceed 20\% of Jupiter's mass, resulting in a dilute core that extends only up to Jupiter's inner $\sim 60~M_{\oplus}$. This leads to a formation scenario that is consistent for both Jupiter and Saturn \citep[see][]{2022arXiv220504100G}, because Saturn is likely to harbour a dilute core that extends to $52-60~M_{\oplus}$ of the planet's total mass \citep{2021NatAs...5.1103M}.

Concerning models with a modification of the EOS, we find a few models with very extended dilute cores but the majority yield $m_{\rm dilute}<0.15$. We suspect that these results are spurious. The changes in the H-He EOS lead to an increase in density at high pressures that can mimic the effect of a dilute core. As the H-He mixture is denser in the dilute core region, less heavy elements can be added, which leads to these low values of $m_{\rm dilute}$ (and also $M_{\rm Z,dil*}$, see Section~\ref{subsec:heavies_distrib}).

\section{Conclusion}

We explored a wide variety of interior models of Jupiter constrained by all available observations and using available EOSs. Our models assume the presence of a central dense compact core, a dilute core of variable extent and heavy-element composition, and an outer envelope of uniform $Z$ composition. The helium phase separation is modelled as a jump in helium abundance in the Mbar pressure range.

While high-pressure experiments and ab initio calculations have led to significant improvements, H-He EOSs remain a source of uncertainty when modelling Jupiter's interior. We observe a range in the adiabatic density profiles of up to 5\% at pressures ranging from 10\,kbar to 10\,Mbar. We interpret these variations as resulting from changes between different EOS tables and from different interpolation choices in relatively sparsely populated tables.

An important source of uncertainty results from our poor knowledge of Jupiter's complex atmosphere and the possibility of a higher entropy than generally assumed. By allowing the $T_{\rm 1\,bar}$ parameter to vary \citep[see][]{2022A&A...662A..18M}, we obtain models that fit all constraints for all EOSs used. The values of $T_{\rm 1\,bar}$ obtained range from 171\,K to 188\,K, significantly higher than 166.1\,K from the Galileo probe \citep{1998JGR...10322815V}, but within $164K-174K,$ which is the range of values obtained by \citet{2022PSJ.....3..159G} from a reanalysis of Voyager's radio occultations. Interestingly, MH13* and HG23+MLS22 (see Table~\ref{tab:eos}), the two EOSs that lead to the lowest $T_{\rm 1\,bar}$ values, yield the highest values of $J_6$, which are slightly outside the range expected from differential rotation. Conversely, the MGF16+MH13 EOS leads to the smallest values of $J_6$, well within expectations, but the largest values of $T_{\rm 1\,bar}$.

In all cases, we obtain a dilute core of heavy-element mass of between $10$ and $25\,M_\oplus$, confirming the result obtained by \citet{2022A&A...662A..18M} that Jupiter's envelope is inhomogeneous. The range of values is also fully compatible with the results obtained by \citet{2022PSJ.....3..185M} and \citet{2019ApJ...872..100D}. The mass of the compact core ranges between 0 and $6\,M_\oplus$. The total mass of heavy elements that we find ranges from 18 to $33\,M_\oplus$.  However, we must stress that, given the possibility of (perhaps significant) superadiabatic regions \citep[see][and Section~\ref{subsec:heavies_distrib}]{Guillot1995, 2012A&A...540A..20L, Leconte+2017}, these masses are lower limits.

Our dilute cores are characterised by a global mass fraction of heavy elements of between 0.02 and 0.27 (in addition to the envelope heavy element mass fraction $Z_1\sim 0.02$) and extend from $\sim$~15\% to $\sim$~60\% of Jupiter's total mass. We reiterate that the exact extent of the dilute core will depend on the shape of its compositional gradient. These solutions therefore encompass those of \citet{2019ApJ...872..100D} and \citet{2022PSJ.....3..185M}, but also allow for small, dilute cores. These solutions with small, dilute cores are compatible with the formation--evolution models of \citet{2020A&A...638A.121M}, which suggest that the outer 80\% in mass should be fully mixed by convection, leaving a primordial dilute core extending only up to Jupiter's inner $\sim 60~M_{\oplus}$. This result is also promising, in light of interior models for  Saturn, which indicate that a dilute core extends to $52-60~M_{\oplus}$ of the planet's total mass \citep{2021NatAs...5.1103M}. This could lead to a formation scenario that is consistent for both Jupiter and Saturn \citep[see][]{2022arXiv220504100G}.

However, there is an important caveat to consider: As pointed out by \citet{2017GeoRL..44.4649W} and \citet{2019ApJ...872..100D}, and confirmed by all further modelling efforts, interior models of Jupiter constrained by Juno's gravitational moments favour solutions with small values of $Z$ in the outer envelope. While our solutions are calculated by imposing $Z_1=0.02$, this represents a bare minimum, given all spectroscopic constraints (see Fig.~\ref{figure:Zatm}). When imposing higher values of $Z_1$, we find an increasingly more difficult situation, with solutions departing from the constraints on $R_{\rm eq}$ and $J_2$ and/or modifications to the EOSs that were too important and most likely incompatible with the experimental constraints. Interestingly, a similar situation arises for Saturn \citep{2021NatAs...5.1103M}, indicating that we may be missing an important physical ingredient.

Several directions of research could lead to significant improvements in our understanding of Jupiter's interior structure and composition. For example, progress in the analysis of Juno microwave radiometer data to infer abundances of ammonia and water as well as temperatures as a function of depth and altitude \citep{2020NatAs...4..609L} will help us to understand heat transport and the composition of the deep atmosphere \citep[see also][]{2022PSJ.....3...74S}. Future radio occultations with Juno should further test observational constraints on temperature and shape. Improvements on EOSs, both experimentally and numerically, with particular emphasis on the hydrogen--helium mixture at pressures between 10~kbar and 10~Mbar and near Jupiter's adiabat (temperatures from 1000~K to 20,000~K on this pressure range) would be extremely valuable. Finally, while indications of the presence of normal modes of Jupiter exist \citep{2011A&A...531A.104G,2022NatCo..13.4632D}, their identification from dedicated observational efforts \citep{Goncalves+2019, Shaw+2022} would be an extremely powerful tool for fully constraining the interior structure and composition.

\begin{acknowledgements}
The authors thank the Juno Interior Working Group for useful discussions and comments. This research was carried out at the Observatoire de la Côte d’Azur under the sponsorship of the Centre National d’Etudes Spatiales.
\end{acknowledgements}

\bibliographystyle{aa} 
\bibliography{aanda}

\begin{appendix} 
\section{Priors in MCMC simulations}
\label{appendix:priors_mcmc}

Table~\ref{tab:priors} lists the priors used for the parameters of our MCMC calculations.

\begin{table*}
\centering
\caption{Parameters explored in our MCMC calculations for dilute core models.}
\begin{tabular}{@{}lcccccc@{}}
\hline
Parameter &  Distribution & Lower bound & Upper bound & $\mu$ & $\sigma$\\
\hline
\hline
$M_{\rm core}$ (M$_{\rm \oplus}$)  &Uniform & 0 & 24     & -- & -- \\
$P_{\rm He}$ (Mbar) &Normal & 0.8 & 9 & 3 & 0.5 \\
$T_{\rm jump}$ (K) & Uniform& 0 & 2000 & -- & -- \\
$Z_{\rm dilute}^{\rm rock}$ &Uniform & 0 & 0.5     & -- & -- \\
$Z_{\rm dilute}^{\rm ice}$  &Uniform & 0 & 0.5     & -- & -- \\
$m_{\rm dilute}$  &Uniform & 0 & 0.6     & -- & -- \\
$T_{\rm 1bar}$ (K) &Normal  & 135 & 215    & 165 & 4 \\
$P_{\rm modif}\, \rm (dyn.cm^{-2})$ & Uniform& $10^{11.5}$ & $10^{12.5}$ & -- & -- \\
$\Delta P$ & Uniform& 0.2 & 0.8 & -- & -- \\
$d\rho$ & Uniform& -0.1 & 0.1 & -- & -- \\
\hline
\end{tabular}
\label{tab:priors}
\begin{flushleft}
\end{flushleft}
\tablefoot{The parameter is given in the first column, the corresponding distribution in the second, and the lower and upper bounds in the third and fourth. When relevant, the mean and the standard deviation of the truncated normal are given in columns five and six. The prior on $T_{\rm 1bar}$ is used in models where $T_{\rm 1bar}$ is not fixed. The priors on $P_{\rm modif}$, $\Delta P,$ and $d\rho$ are used in models allowing for a modification of the EOS. We recall that $Y_{\rm atm}=0.238$ and $Y_{\rm proto}=0.277$.}
\end{table*}

\section{Comparison between runs with Gaussian and uniform priors}
\label{appendix:unif_vs_gauss}

\begin{figure*}
   \resizebox{\hsize}{!}
            {\includegraphics{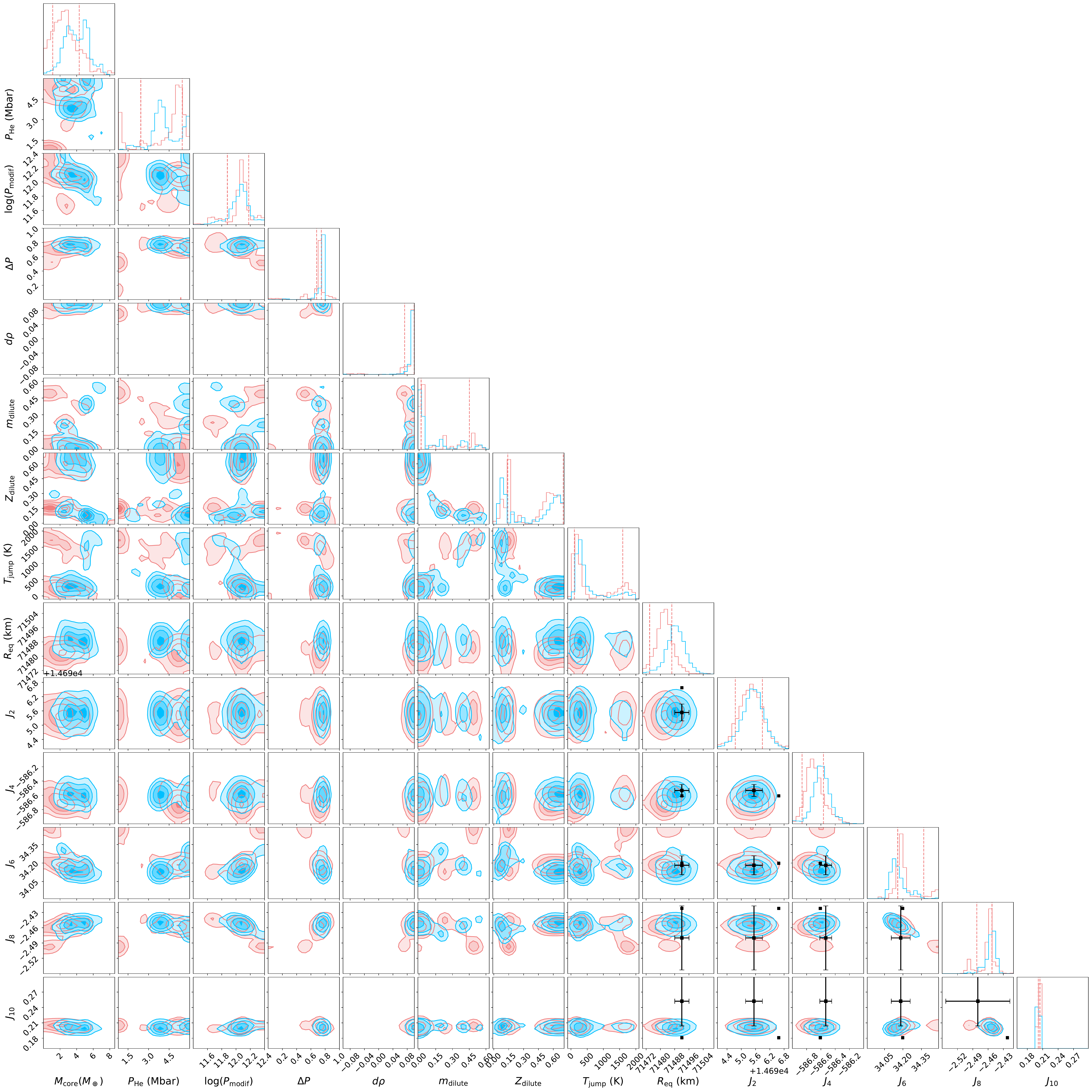}}
      \caption{Posterior distributions obtained with a modification of the EOS. Red shows results obtained with Gaussian priors. Blue shows results obtained with uniform priors. $T_{\rm 1bar}$ is fixed at 166.1~K and $Z_{1}=0.02$. The black points correspond to the measured $J_{2n}$ by Juno. The black error bars correspond to Juno's measurements accounting for differential rotation for the $J_{2n}$.
              }
         \label{appendix:cornerplot_unifvsgauss}
\end{figure*}

Figure~\ref{appendix:cornerplot_unifvsgauss} compares the posterior distributions of two MCMC simulations: using Gaussian or uniform priors on the parameters to modify the EOS (see Section~\ref{subsec:runs_modif}).

\section{Corner plots of models with and without modification of the EOS} \label{appendix:cornerplots}

\begin{figure*}
   \resizebox{\hsize}{!}
            {\includegraphics{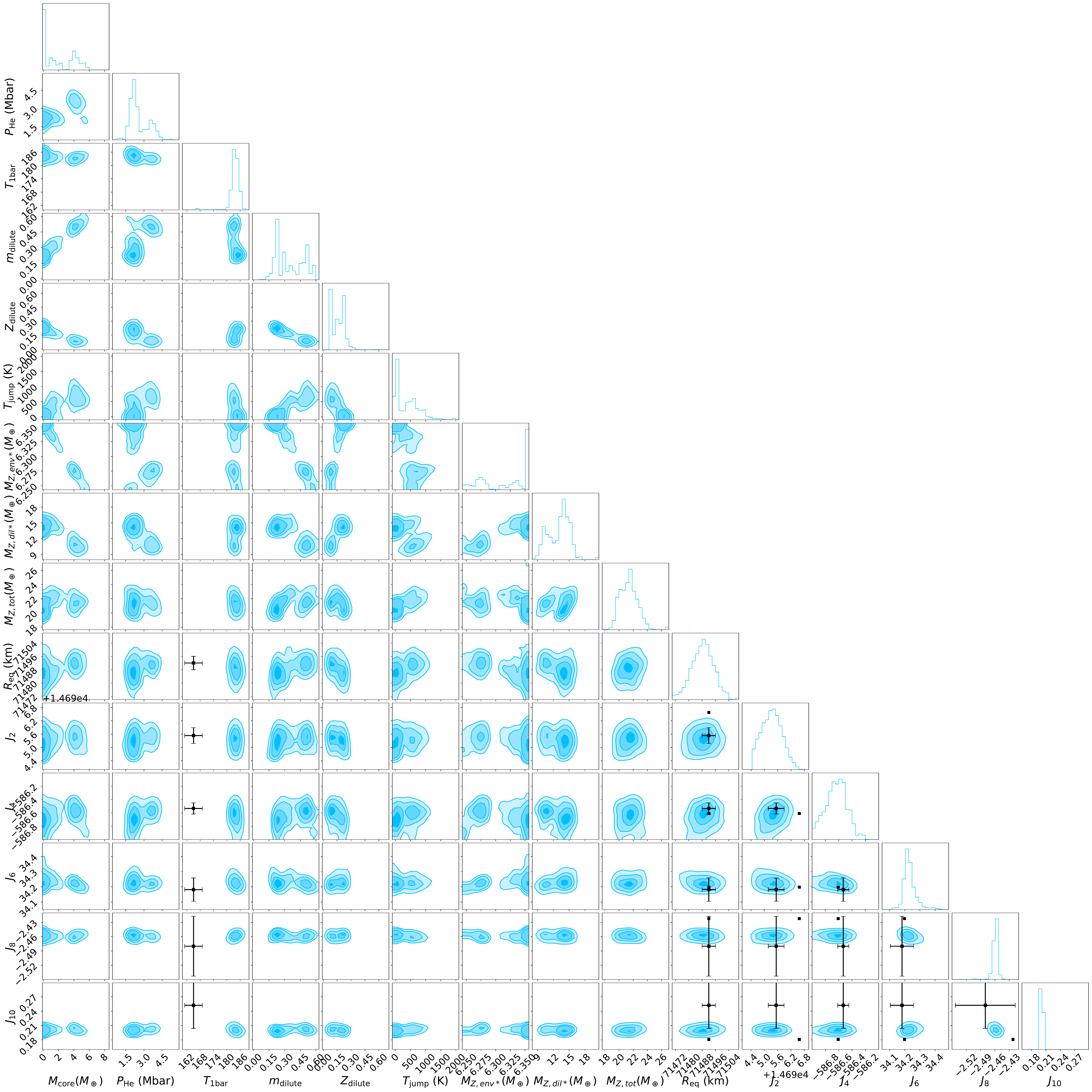}}
      \caption{Posterior distributions obtained with the MGF16+MH13 EOS, where $T_{\rm 1bar}$ is a free parameter, $Z_{1}=0.02$ ($1.3~\times$ the protosolar value). The black points correspond to the measured $J_{2n}$ by Juno. The black error bars correspond to Juno's measurements accounting for differential rotation for the $J_{2n}$ and Galileo's measurement for $T_{\rm 1bar}$.
              }
         \label{appendix:NOmodifEOS_MH13}
\end{figure*}

\begin{figure*}
   \resizebox{\hsize}{!}
            {\includegraphics{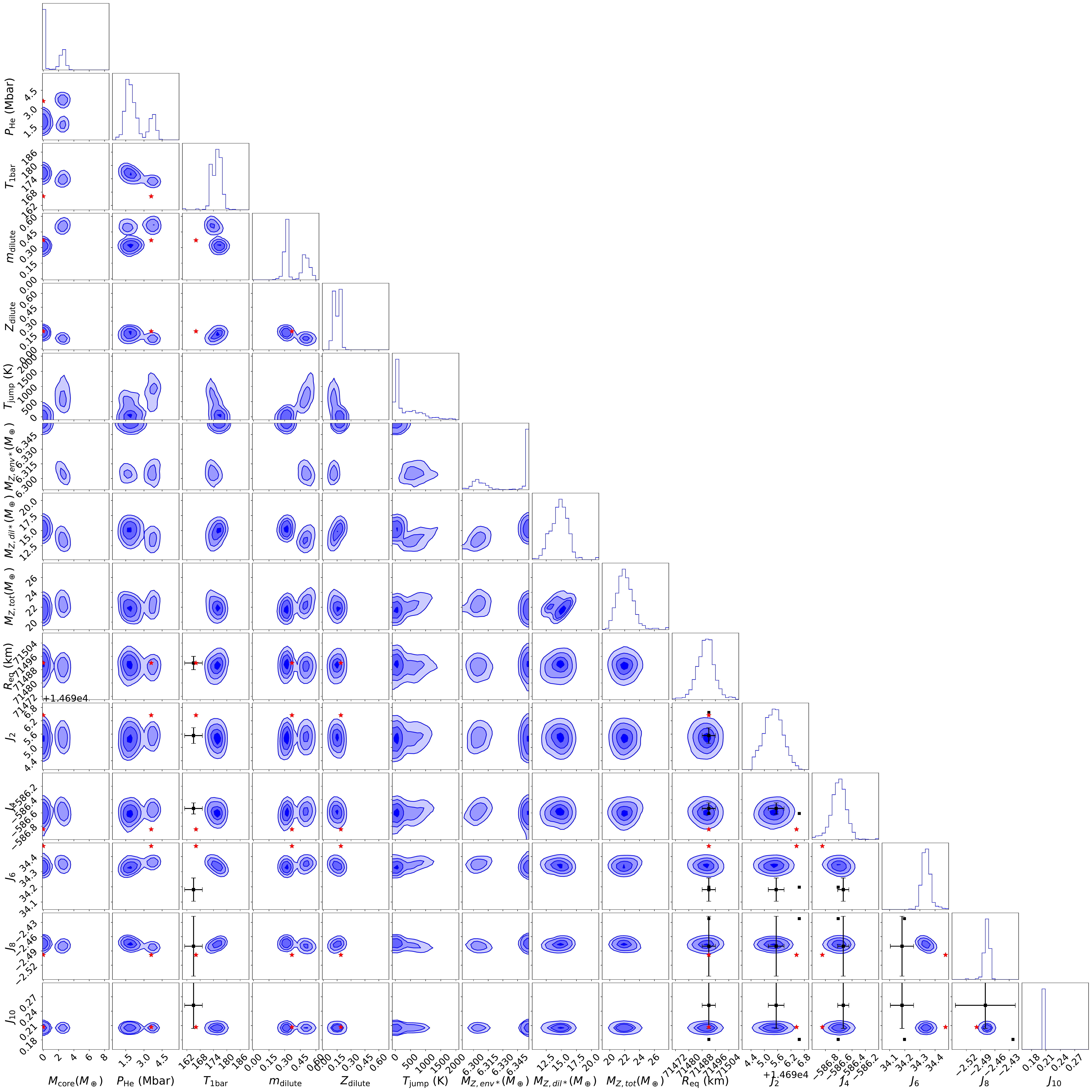}}
      \caption{Same as Fig.~\ref{appendix:NOmodifEOS_MH13} but with the MH13* EOS. The red star shows the \citet{2022PSJ.....3..185M} preferred (static) model.
              }
         \label{appendix:NOmodifEOS_MH13star}
\end{figure*}

\begin{figure*}
   \resizebox{\hsize}{!}
            {\includegraphics{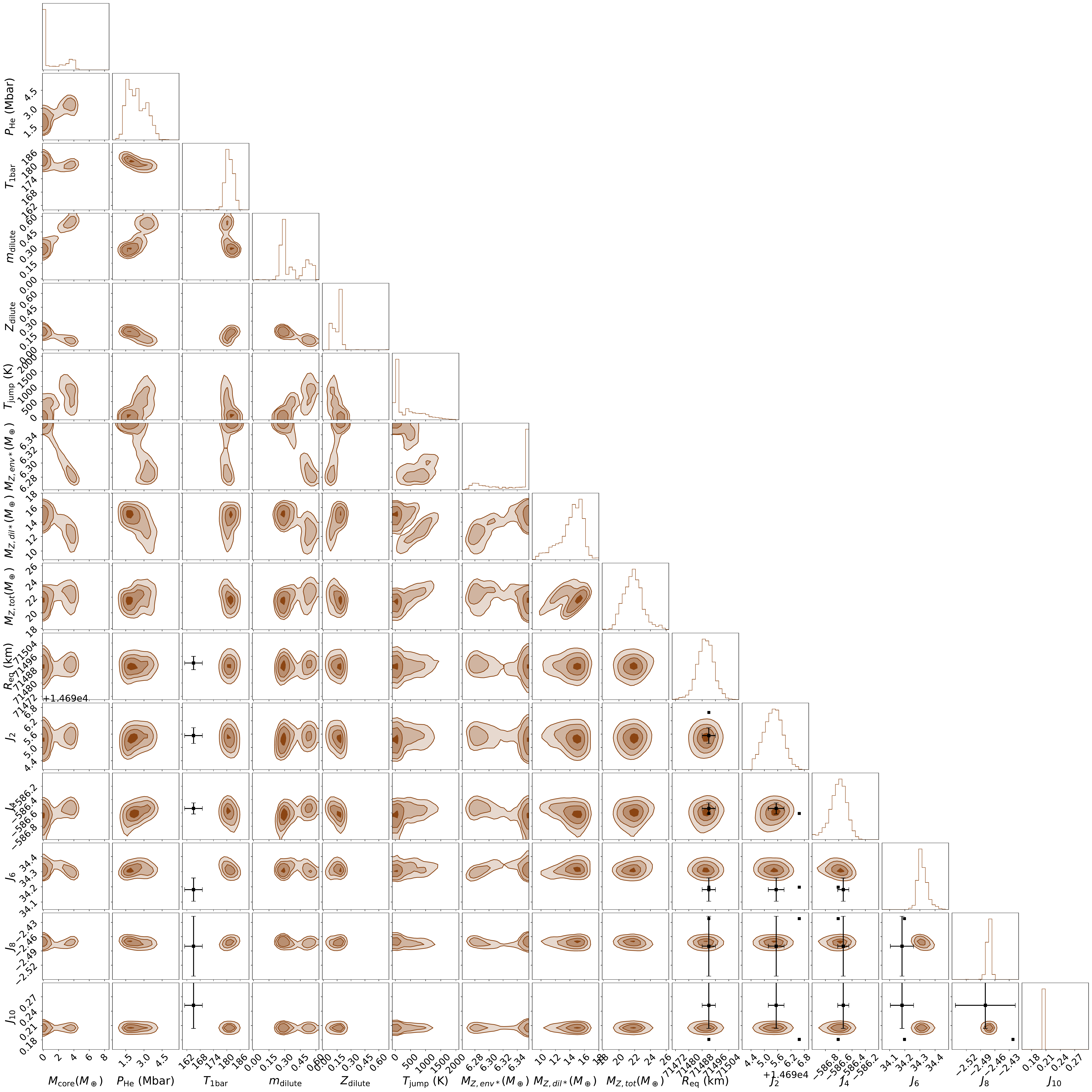}}
      \caption{Same as Fig.~\ref{appendix:NOmodifEOS_MH13} but with the CD21 EOS.
              }
         \label{appendix:NOmodifEOS_CD21}
\end{figure*}

\begin{figure*}
   \resizebox{\hsize}{!}
            {\includegraphics{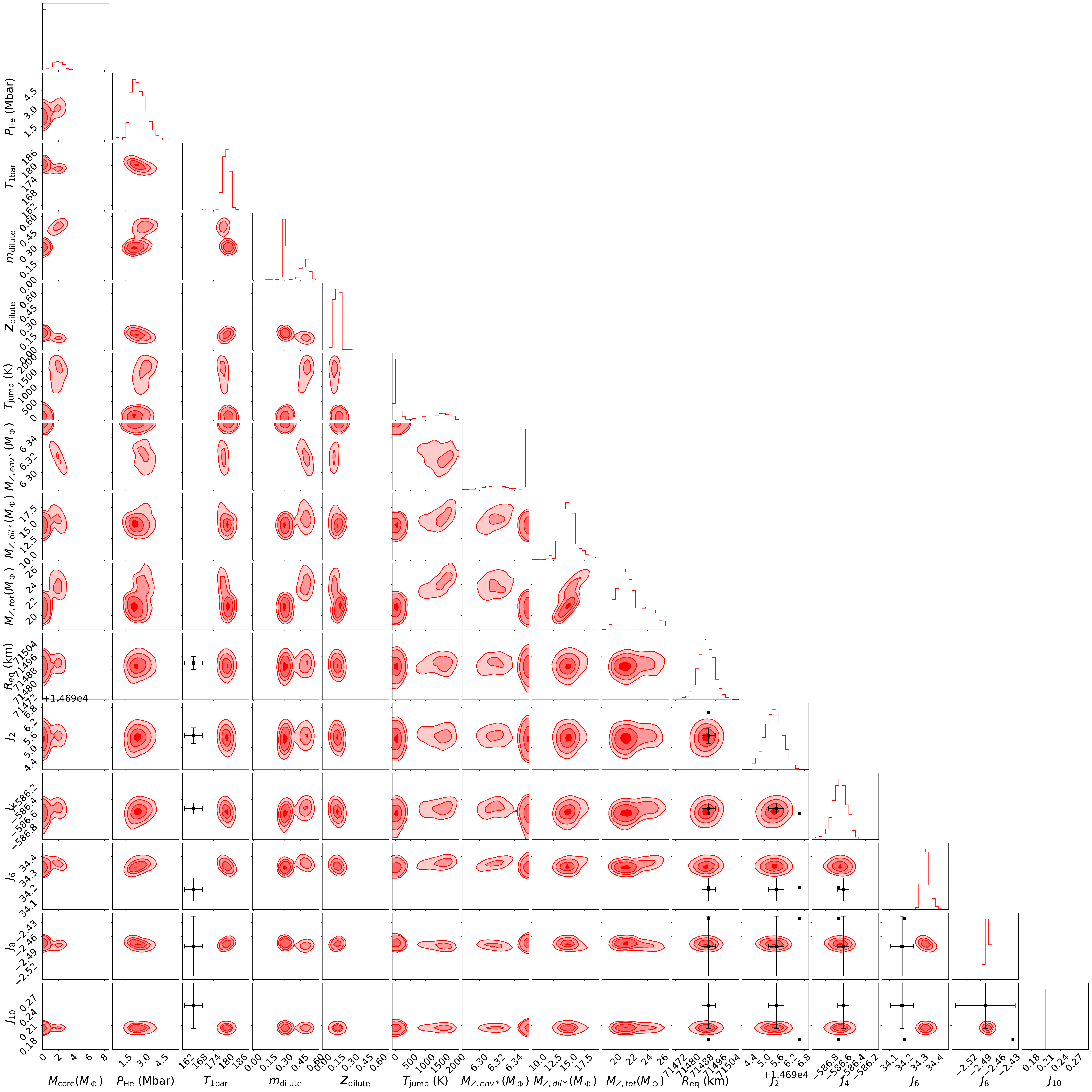}}
      \caption{Same as Fig.~\ref{appendix:NOmodifEOS_MH13} but with the HG23+CMS19 EOS.
              }
         \label{appendix:NOmodifEOS_CMS19wNIE}
\end{figure*}

\begin{figure*}
   \resizebox{\hsize}{!}
            {\includegraphics{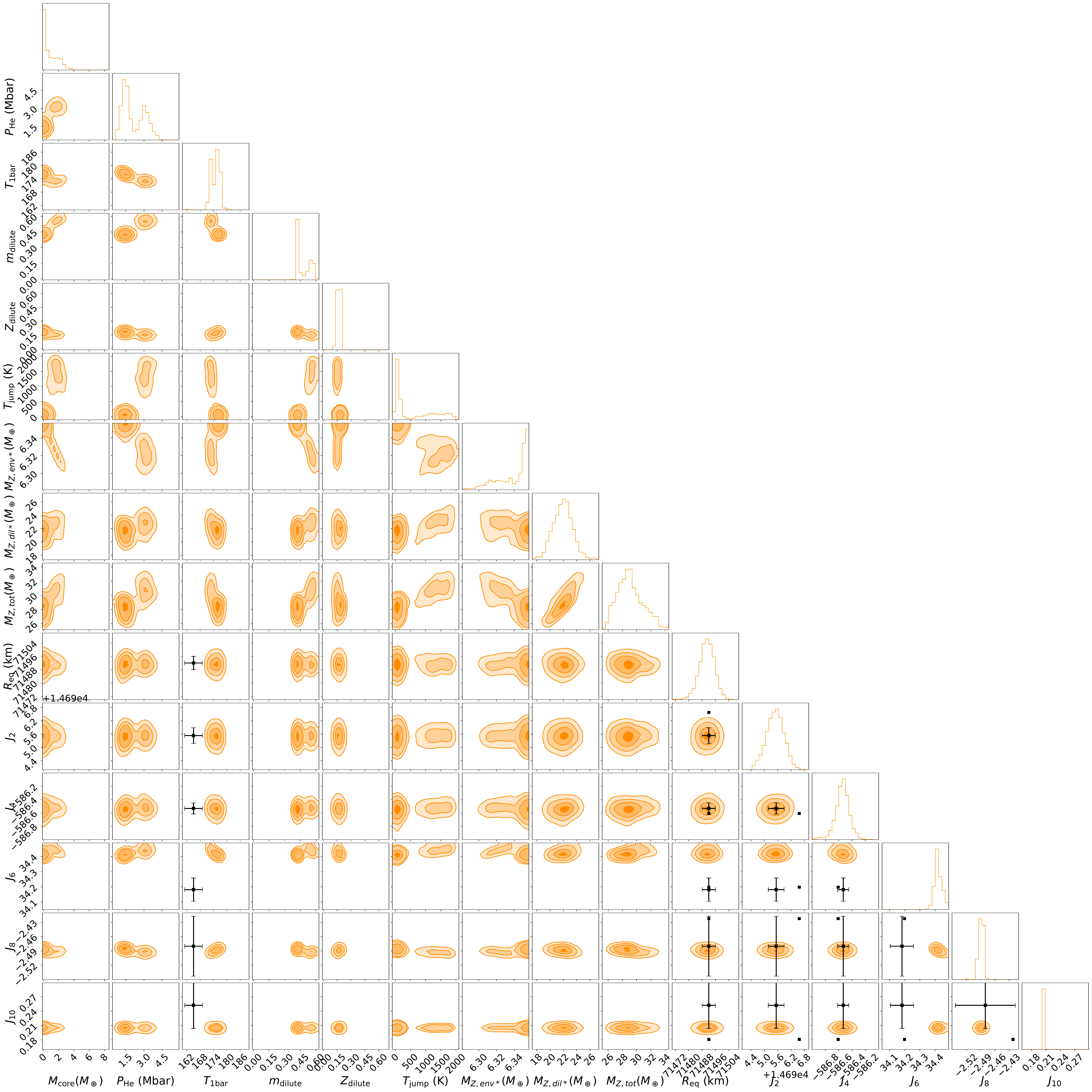}}
      \caption{Same as Fig.~\ref{appendix:NOmodifEOS_MH13} but with the HG23+MLS22 EOS.
              }
         \label{appendix:NOmodifEOS_MLS20wNIE}
\end{figure*}

Figures~\ref{appendix:NOmodifEOS_MH13}, \ref{appendix:NOmodifEOS_MH13star}, \ref{appendix:NOmodifEOS_CD21}, \ref{appendix:NOmodifEOS_CMS19wNIE}, and \ref{appendix:NOmodifEOS_MLS20wNIE} show the posterior distributions of the MCMC simulations using original EOSs, respectively, MGF16+MH13, MH13*, CD21, HG23+CMS19, and HG23+MLS22.

\begin{figure*}
   \resizebox{\hsize}{!}
            {\includegraphics{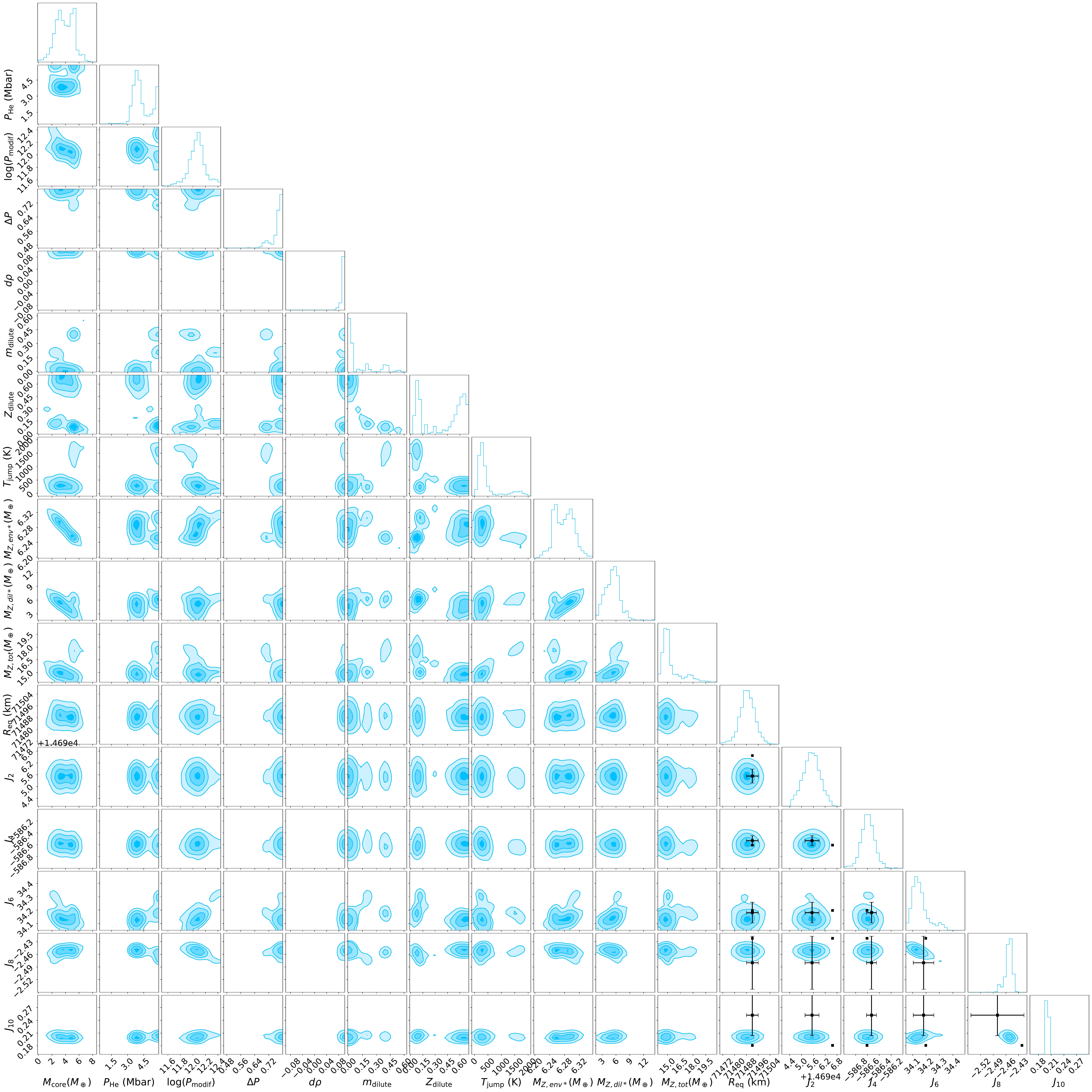}}
      \caption{Posterior distributions obtained with a modification of the EOS, where $T_{\rm 1bar}$ is fixed at 166.1~K, $Z_{1}=0.02$ ($1.3~\times$ the protosolar value). The black points correspond to the measured $J_{2n}$ by Juno. The black error bars correspond to Juno's measurements accounting for differential rotation for the $J_{2n}$.
              }
         \label{appendix:modifEOS_MH13}
\end{figure*}

\begin{figure*}
   \resizebox{\hsize}{!}
            {\includegraphics{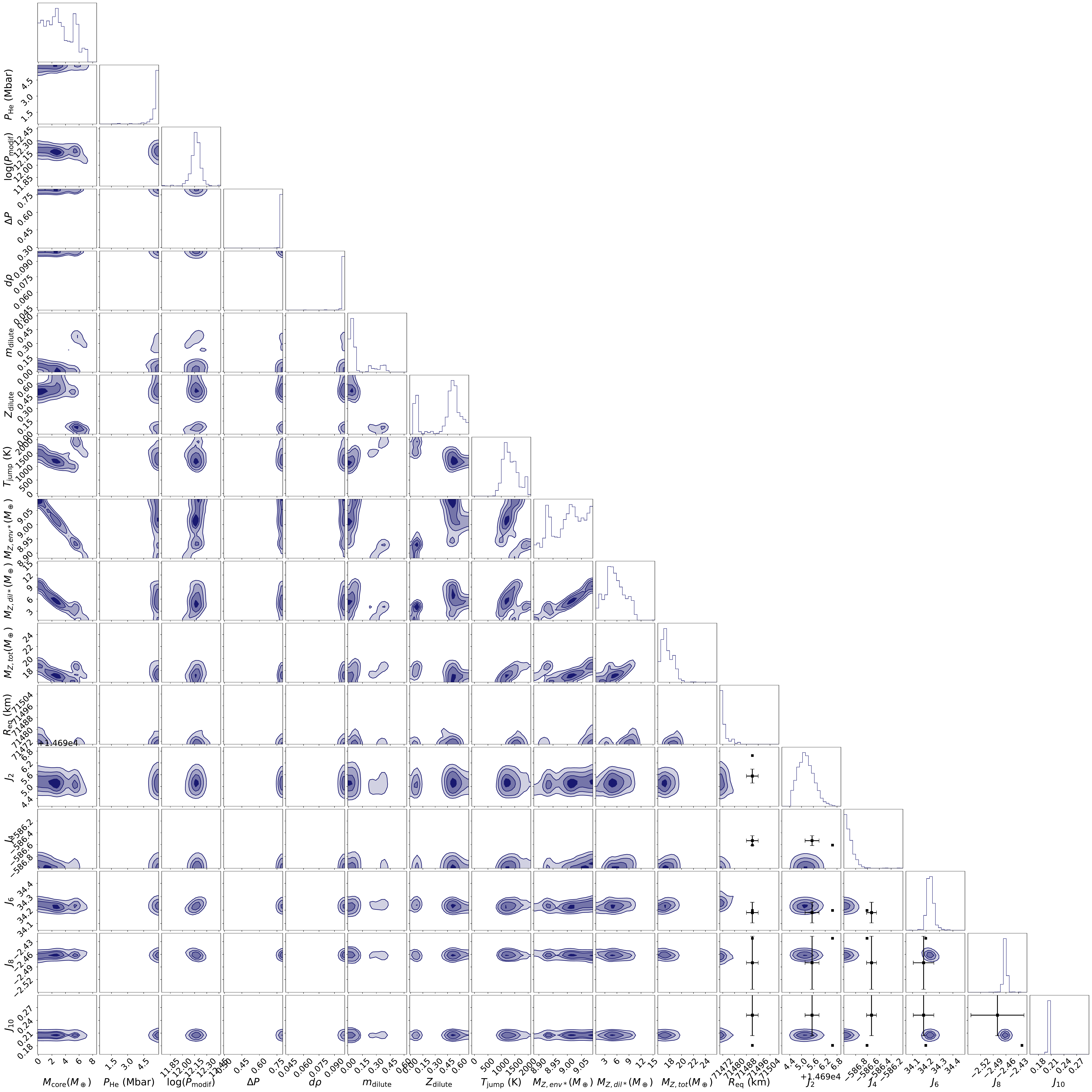}}
      \caption{Same as Fig.~\ref{appendix:modifEOS_MH13} but with $Z_{1}=0.0286$ ($1.9~\times$ the protosolar value)
              }
         \label{appendix:modifEOS_MH13_Z00286}
\end{figure*}

\begin{figure*}
   \resizebox{\hsize}{!}
            {\includegraphics{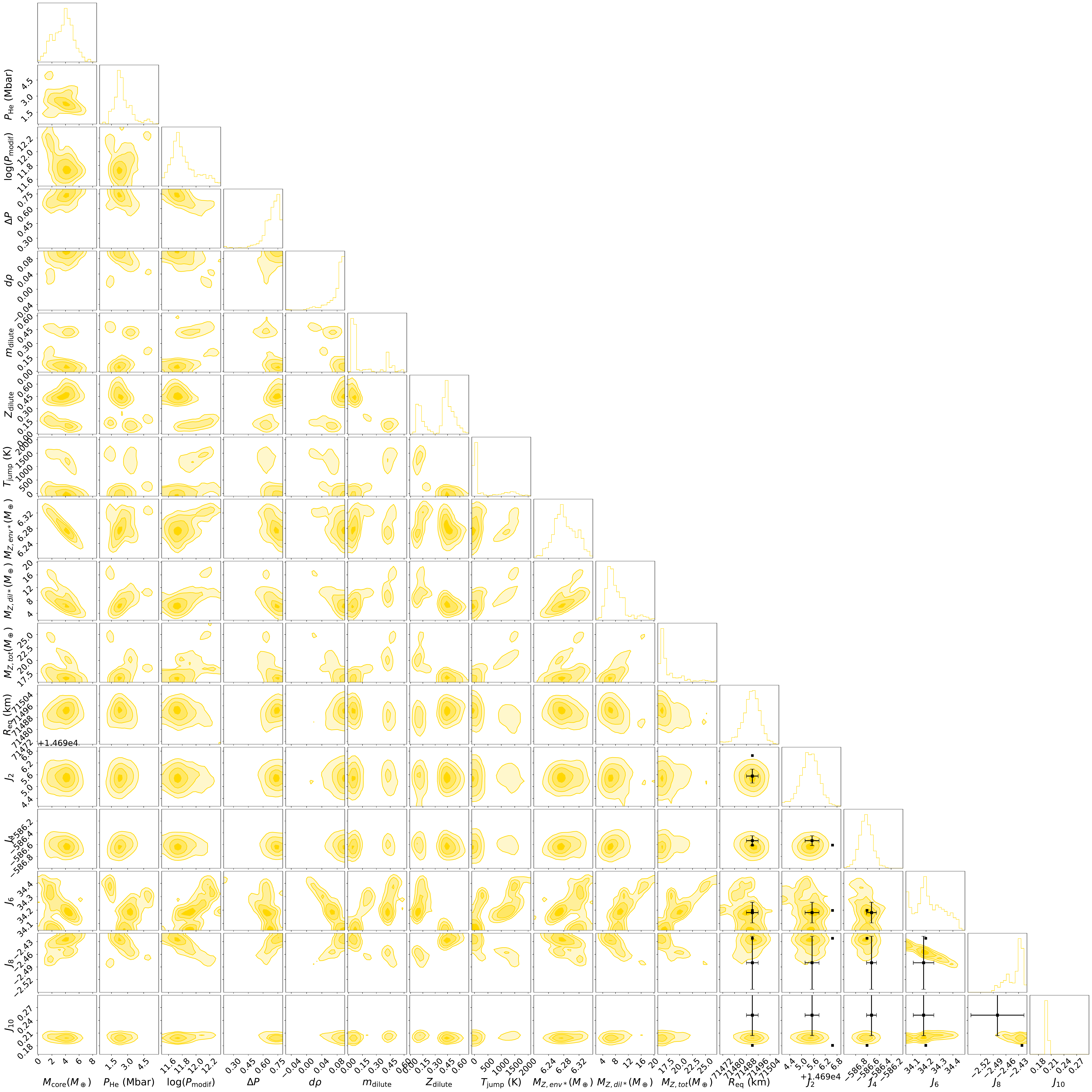}}
      \caption{Posterior distributions obtained with a modification of the EOS, where $T_{\rm 1bar}$ is fixed at 174.1~K, $Z_{1}=0.02$ ($1.3~\times$ the protosolar value). The black points correspond to the measured $J_{2n}$ by Juno. The black error bars correspond to Juno's measurements accounting for differential rotation for the $J_{2n}$.
              }
         \label{appendix:modifEOS_MH13_Z002_T174}
\end{figure*}

\begin{figure*}
   \resizebox{\hsize}{!}
            {\includegraphics{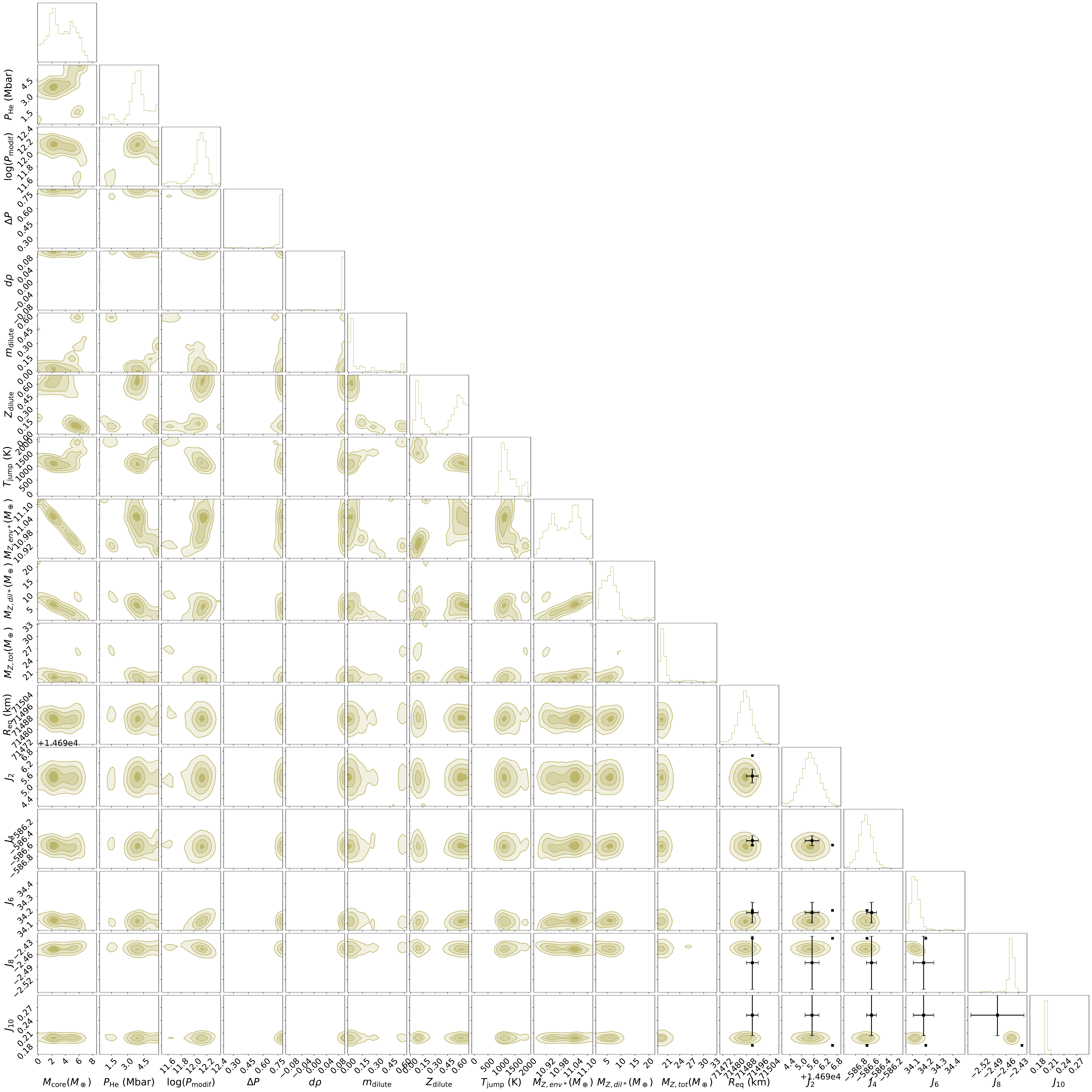}}
      \caption{Same as Fig.~\ref{appendix:modifEOS_MH13_Z002_T174} but with $Z_{1}=0.035$ ($2.3~\times$ the protosolar value)
              }
         \label{appendix:modifEOS_MH13_Z0035_T174}
\end{figure*}

Figures~\ref{appendix:modifEOS_MH13} and \ref{appendix:modifEOS_MH13_Z00286} show the posterior distributions of the MCMC simulations using modified EOS, with $T_{\rm 1bar}=166.1$~K and respectively for $Z_1=0.02$ and $Z_1=0.0286$.
Figures~\ref{appendix:modifEOS_MH13_Z002_T174} and \ref{appendix:modifEOS_MH13_Z0035_T174} show the posterior distributions of the MCMC simulations using modified EOS, with $T_{\rm 1bar}=174.1$~K and respectively for $Z_1=0.02$ and $Z_1=0.035$.

\section{Comparison with \citet{2022PSJ.....3..185M}}
\label{appendix:comp_BM}

\begin{figure*}
   \resizebox{\hsize}{!}
            {\includegraphics{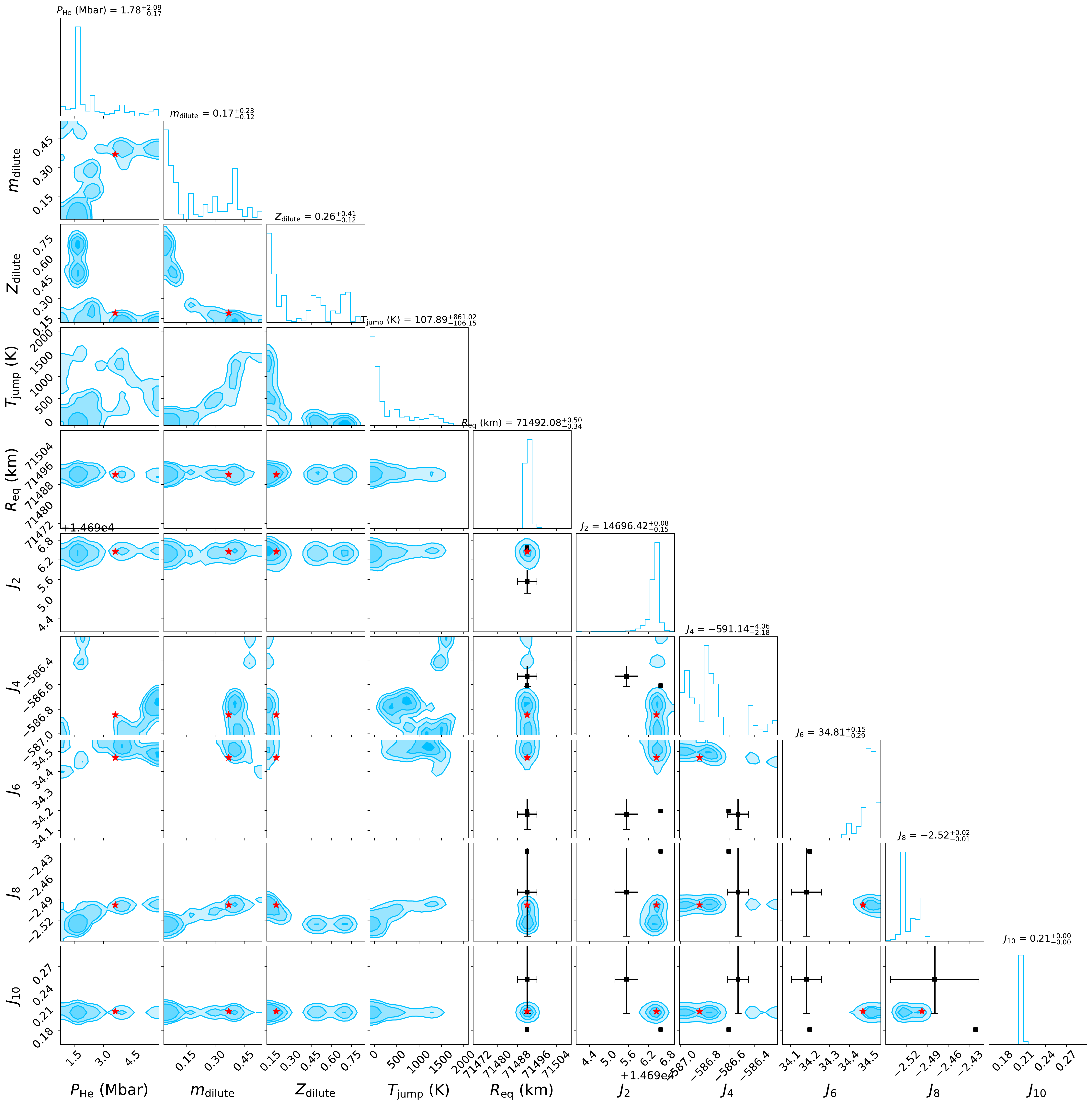}}
      \caption{Posterior distributions obtained with MH13*. The red star shows the \citet{2022PSJ.....3..185M} preferred (static) model. $M_{\rm core}=0$, $T_{\rm 1bar}$ is fixed at 166.1~K and $Z_{1}=0.0153$. The black error bars correspond to Juno's measurements accounting for differential rotation for the $J_{2n}$.
              }
         \label{appendix:comp_BM_figure}
\end{figure*}

We ran MCMC simulations to reproduce the results obtained by \citet{2022PSJ.....3..185M}. To do so, we changed the values of the gravitational moments around which the MCMC is sampling models. We used the gravitational moments of the interior model of \citet{2022PSJ.....3..185M} (see their Table 1). We used the same properties: no compact core, $Z_1=0.0153$, $T_{\rm 1bar}=166.1$~K, and the MH13* EOS. Figure~\ref{appendix:comp_BM_figure} shows the posterior distributions we obtain. We find models with similar properties to Model A from \citet{2022PSJ.....3..185M}: the same gravitational moments, the same $P_{\rm He}$, and comparable characteristics for the dilute core.

\section{Subsample of models}
We extracted one single model from each MCMC simulation using original EOSs (see Fig.~\ref{appendix:NOmodifEOS_MH13}, \ref{appendix:NOmodifEOS_MH13star}, \ref{appendix:NOmodifEOS_CD21}, \ref{appendix:NOmodifEOS_CMS19wNIE}, \ref{appendix:NOmodifEOS_MLS20wNIE}) and list them in Table~\ref{tab:bestmodels1} and Table~\ref{tab:bestmodels2}.

\begin{table*}
\centering
\caption{Comparison of the parameters of selected models extracted from MCMC simulations using original EOSs.}
\begin{tabular}{@{}lcccccccccccc@{}}
\hline
EOS & $M_{\rm core}$ (M$_{\rm \oplus}$) & $P_{\rm He}$ (Mbar) & $m_{\rm dilute}$ & $Z_{\rm dilute}$ & $T_{\rm jump}$ (K) & $T_{\rm 1bar}$ (K) & $M_{\rm Z,env*}$ (M$_{\rm \oplus}$) & $M_{\rm Z,dil*}$ (M$_{\rm \oplus}$) & $M_{\rm Z,tot}$ (M$_{\rm \oplus}$) \\
\hline
\hline
MGF16+MH13 & 0.0305 & 2.12 & 0.237 & 0.214 & 60.3 & 185.3 & 6.52 & 14.46 & 20.98 \\
MH13* & 0.0288 & 2.05 & 0.317 & 0.175 & 82.3 & 176.1 & 6.52 & 15.51 & 22.03 \\
CD21 & 0.0115 & 1.64 & 0.280 & 0.186 & 16.1 & 183.3 & 6.52 & 14.57 & 21.08 \\
HG23+CMS19 & 0.231 & 1.66 & 0.324 & 0.177 & 9.10 & 182.1 & 6.52 & 16.05 & 22.57 \\
HG23+MLS22 & 0.466 & 2.06 & 0.441 & 0.180 & 60.0 & 174.6 & 6.52 & 22.19 & 28.71 \\
\hline
\end{tabular}
\label{tab:bestmodels1}
\begin{flushleft}
\end{flushleft}
\tablefoot{Models are available at the CDS via anonymous ftp to \href{http://cdsarc.cds.unistra.fr}{cdsarc.cds.unistra.fr} (\href{ftp://130.79.128.5/}{130.79.128.5})
or via \url{https://cdsarc.cds.unistra.fr/cgi-bin/qcat?J/A+A/} and at \url{https://doi.org/10.5281/zenodo.7598377}.}
\end{table*}

\begin{table*}
\centering
\caption{Comparison of the equatorial radius and the gravitational moments of the same models as in Table~\ref{tab:bestmodels1}.}
\begin{tabular}{@{}lcccccccccccc@{}}
\hline
EOS & $R_{\rm eq}$ (km )& $J_2 \times 10^6$ & $J_4 \times 10^6$ & $J_6 \times 10^6$ & $J_8 \times 10^6$ & $J_{10} \times 10^6$\\
\hline
\hline
MGF16+MH13 & 71487.6 & 14695.42 & -586.649 & 34.211 & -2.4548 & 0.2011\\
MH13* & 71492.1 & 14695.62 & -586.622 & 34.339 & -2.4750 & 0.2035 \\
CD21 & 71491.8 & 14695.57 & -586.611 & 34.291 & -2.4676 & 0.2027 \\
HG23+CMS19 & 71491.4 & 14695.53 & -586.559 & 34.309 & -2.4704 & 0.2030 \\
HG23+MLS22 & 71491.2 & 14695.65 & -586.625 & 34.436 & -2.4904 & 0.2054\\
\hline
\end{tabular}
\label{tab:bestmodels2}
\begin{flushleft}
\end{flushleft}
\end{table*}

\end{appendix}

%
   
%

\end{document}